\documentclass[lettersize,journal]{IEEEtran}
\usepackage{amsmath,amsfonts}
\usepackage{algorithmic}
\usepackage{algorithm}
\usepackage{array}
\usepackage[caption=false,font=normalsize,labelfont=sf,textfont=sf]{subfig}
\usepackage{textcomp}
\usepackage{stfloats}
\usepackage{url}
\usepackage{verbatim}
\usepackage{graphicx}
\usepackage{cite}

\usepackage{multicol,multirow}
\usepackage{xcolor}

\DeclareGraphicsExtensions{.pdf,.png,.jpg}
\graphicspath{{images/}{postproc_afterpulse/}}

\begin{document}
\title{Dead time duration and active reset influence on the afterpulse probability of InGaAs/InP single-photon avalanche diodes}

\author{A.~V.~Losev${}^{1,2,3,4}$, V.~V.~Zavodilenko${}^{1,4}$, A.~A.~Koziy${}^{1,4}$, A.~A.~Filyaev${}^{1,4,5}$, K.~I.~Khomyakova${}^{6}$, Y.~V.~Kurochkin${}^{1,3,4}$, A.~A.~Gorbatsevich${}^{2,7}$\\
{${}^1$"QRate" LLC, St. Novaya, d. 100, Moscow region, Odintsovo, Skolkovo, 143026, Russia. \\
${}^2$National Research University of Electronic Technology MIET, Shokin Square, 1, Zelenograd, 124498, Russia. \\
${}^3$National University of Science and Technology MISIS, Leninsky prospect, 4, Moscow, 119333, Russia. \\
${}^4$NTI Center for Quantum Communications, National University of Science and Technology MISiS, Leninsky prospekt 4, Moscow, 119049, Russia. \\
${}^5$Bauman Moscow State Technical University, 2nd Baumanskaya st., 5, Moscow, 105005, Russia. \\
${}^6$National Research Tomsk State University, Lenin Av. 36, Tomsk, 634050, Russia. \\
${}^7$P.N. Lebedev Physical Institute of the Russian Academy of Sciences,  Leninsky prospect, 53, Moscow, 119333, Russia.}} 

\maketitle

\begin{abstract}
	We have performed a detailed study of the dependence of afterpulse probability in InGaAs/InP sine-gated SPAD on the dead time and the existing approach for its implementation. We demonstrated an electrical scheme combining sinusoidal gating and active reset. We have shown when such solutions are beneficial from the key distribution point of view and enough to use a simple scheme with a snatching comparator. We have also proposed a precise method for measuring the afterpulse and presented a model describing the non-markovian dynamic of this effect. We have demonstrated that our afterpulsing measurement approach makes these measurements less dependent on the parameters of the flow of the diode and the laser pulses power changes, in contrast to the other methods considered in this paper.
\end{abstract}


\section{Introduction}\label{sec:intro}

\IEEEPARstart{D}{etectors} based on superconducting nanowires (SNSPD) \cite{you2020superconducting} and  single photon avalanche diodes (SPAD) \cite{bruschini2017ten} have proven themselves in the best way as single-photon detectors (SPDs). Each of the implementations has both its advantages and disadvantages. SNSPD has high probability of photon detection and low noise level, but it is large and quite expensive due to usage of helium cryostat \cite{chang2021detecting}. SPAD-based SPDs have small size and low cost, but detection probability is relatively low, and noise characteristics are high. In quantum key distribution (QKD), both first and second types of SPD have found their application \cite{agnesi2020simple, zhang2018experimental}. It is advisable to use SNSPD for key distribution over long distances, both over fiber and open space. SNSPD based QKD was able to demonstrate key distribution distance records \cite{chen2020sending}. It is advisable to use SPAD-based SPD in small-sized industrial installations \cite{kiktenko2017demonstration} that distribute the key within one city or even one building since the loss of photons in the line is minimal and the key generation rate is relatively high.

One of the big problems of the QKD is the difficulty of determining the secrecy of the generated key. In contrast to classical cryptographic algorithms, in which applied mathematical transformations strictly determine  confidentiality of the key, quantum cryptography depends on the installation's physical parameters \cite{zhao2021practical}. Thus, calculated key secrecy can be selected pessimistically, which will significantly reduce key generation rate or optimistically, which will endanger security of subsequently encrypted data. For this reason, development of methods for accurate determination of the parameters of SPD is an actual task that can significantly increase the efficiency of the QKD installation as a whole \cite{wang2019afterpulsing}.

In SNSPD with adequate control electronics, there are no effects associated with previous triggers \cite{wang2019fast}. Therefore, we can consider all processes as Markovian \cite{wein2020analyzing}. There is "memory" of earlier triggers in SPAD-based SPDs -- the processes have more complicated influence on counting statistics \cite{sarbazi2018impact}, and the construction of a global SPD model becomes complicated. Charges captured by the traps cause this memory. These charges relax after a particular time and can lead to the formation of an avalanche and the subsequent detector's triggering, and are called afterpulses \cite{smirnov2018sequences}.

For detectors of $1550$ nm wavelength photons, based on InGaAs/InP SPAD, total relaxation time is about $1 - 50 \ \mu s $ at a temperature of about $-100 - -50 \ {}^\circ$C. For Si-based photon detectors for visible radiation, full relaxation time is about $200$ ns \cite{kramnik2020efficient}. To overcome afterpulsing, the following methods are used: usage of SPD circuits with dead time -- the time during which  the detector is not able to detect photons  after the previous detection event; lowering avalanche growing time by fast quenching.  We can use high dead time, but SPDs used in QKD receive significant restrictions on the limiting operating frequencies, which harms the installation's efficiency as a whole. Therefore, to obtain the SPD's highest efficiency for the QKD application, the dead time value is reduced to $1- 10 \ \mu s$ while sacrificing QBER but gaining maximum count rate, which is especially important for short distances QKD \cite{fan2020optimizing}.

Two articles inspire the theoretical part of our work: \cite{owens1994photon}, and \cite{wang2016non}, where authors try to describe the recurrent nature of the afterpulsing effect and its non-markovian properties. Our article used more rigorous probability equations, making our model more general, for example, for very different count rates. We developed an accurate method for calculating the detector's afterpulse probability from counting statistics.

In this work we observe only the dead time approach for afterpulse control. We investigated two possible technical realizations of the dead time: it includes comparator latching (passive quenching and reset) as well as active bias lowering (passive-active quenching and reset).  We analyzed these two methods and made recommendations on using one of them in different practical cases. We made recommendations for setting time parameters of latching the comparator based on statistical data analysis and the  physics of the processes.

\section{Common (standard) afterpulse models}\label{sec:popular}

\noindent There are the most popular methods for determining of afterpulsing: Bethune method \cite{kang2003afterpulsing, bethune2004high}, Yuan method \cite{yuan2007high, namekata20091, nambu2011efficient}, coincidence method \cite{zhang2009practical, zhang20102, zhang2014electro}, double-pulse method \cite{restelli2012time, itzler2011advances, zhang2009comprehensive}, autocorrelation method \cite{owens1994photon, arahira2016effects},  Klyshko method \cite{Klyshko_1980, kwiat1994absolute, brida2000quantum, polyakov2007high}.  The double-pulse method can be used only for changeable gates, like square gates, and is not applicable for sine-gating, and we do not overview it here. The autocorrelation method can be effectively used only with a multichannel autocorrelator device, and we do not observe this method due to its absence. The Klyshko method was used as a true single-photon source based on a parametric downconversion effect instead of a simple laser with multiphoton states. However, we cannot observe this method because we do not have a true single-photon source. We will briefly describe the essence of each of the other common methods.

\begin{figure}[ht]\centering
	\includegraphics[width=0.47\linewidth]{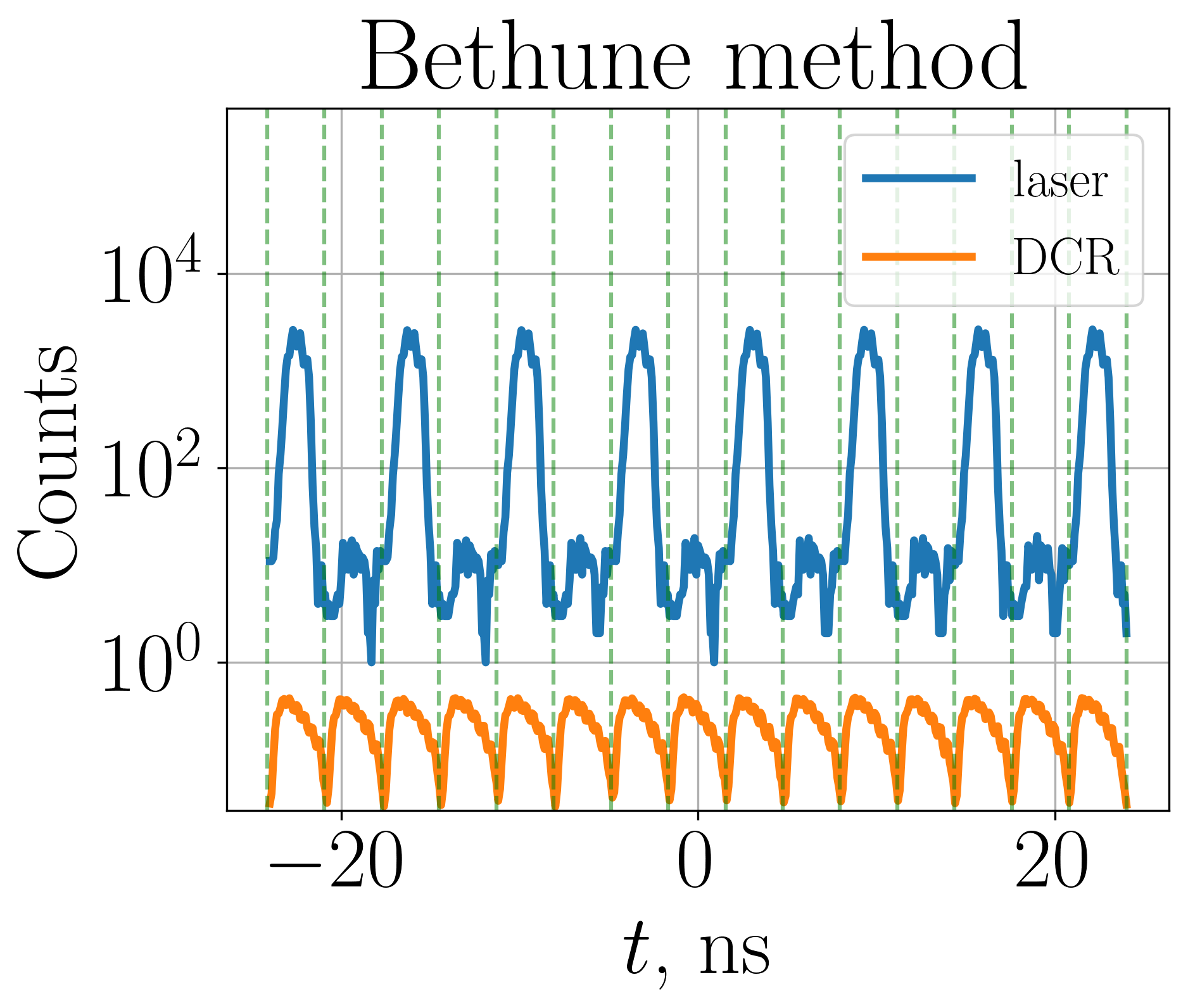}
	\includegraphics[width=0.47\linewidth]{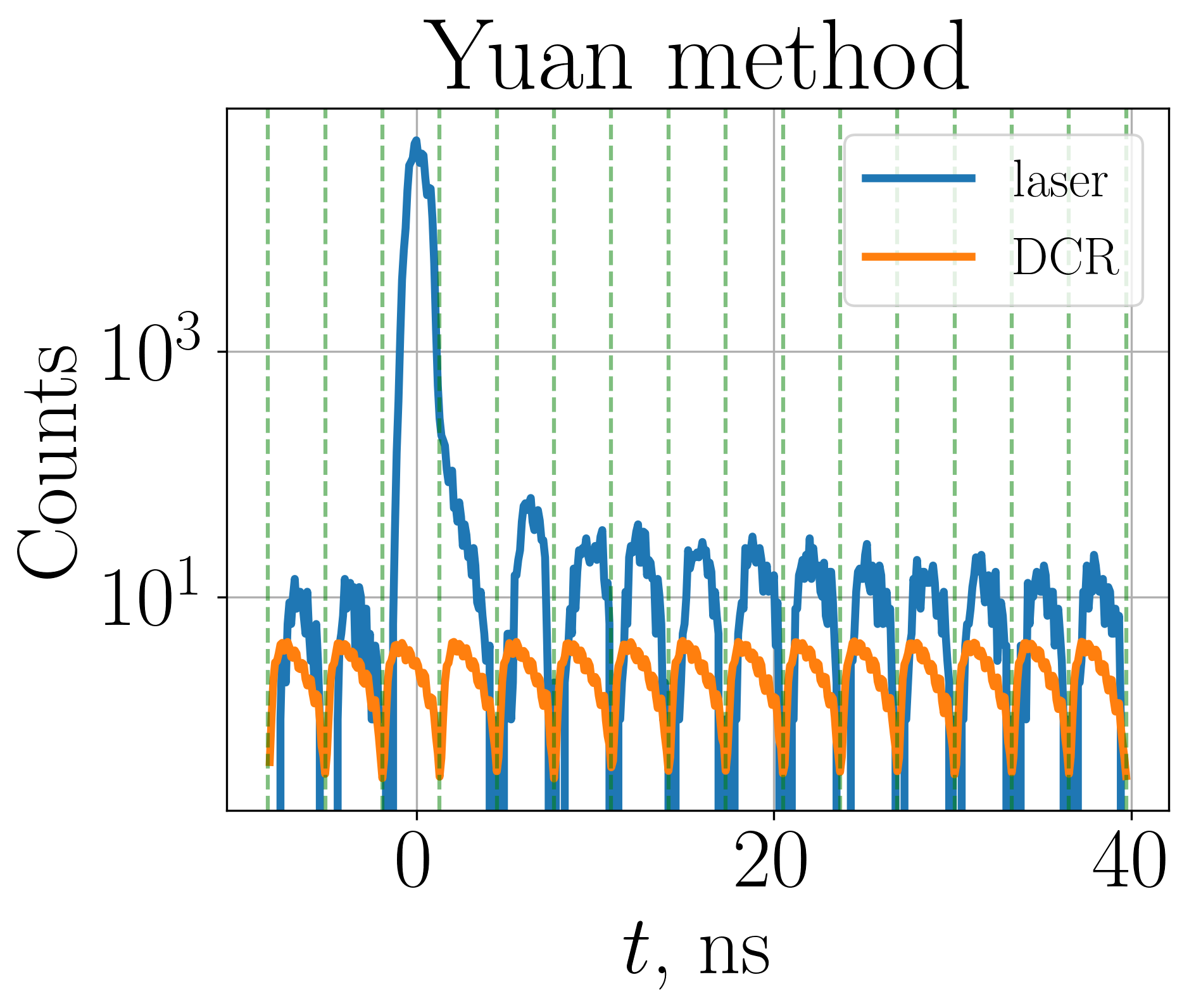}
	\includegraphics[width=0.47\linewidth]{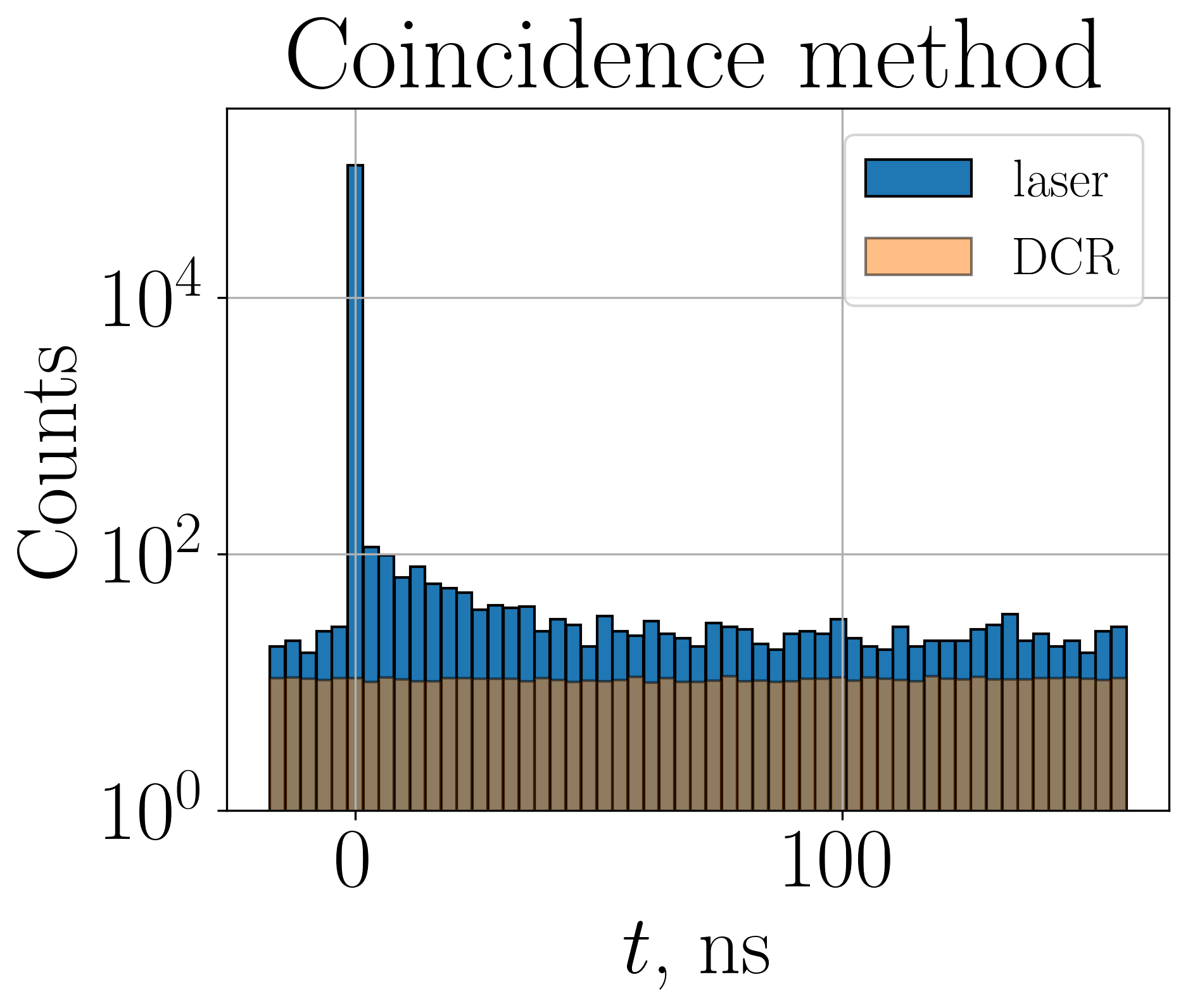}
	\includegraphics[width=0.47\linewidth]{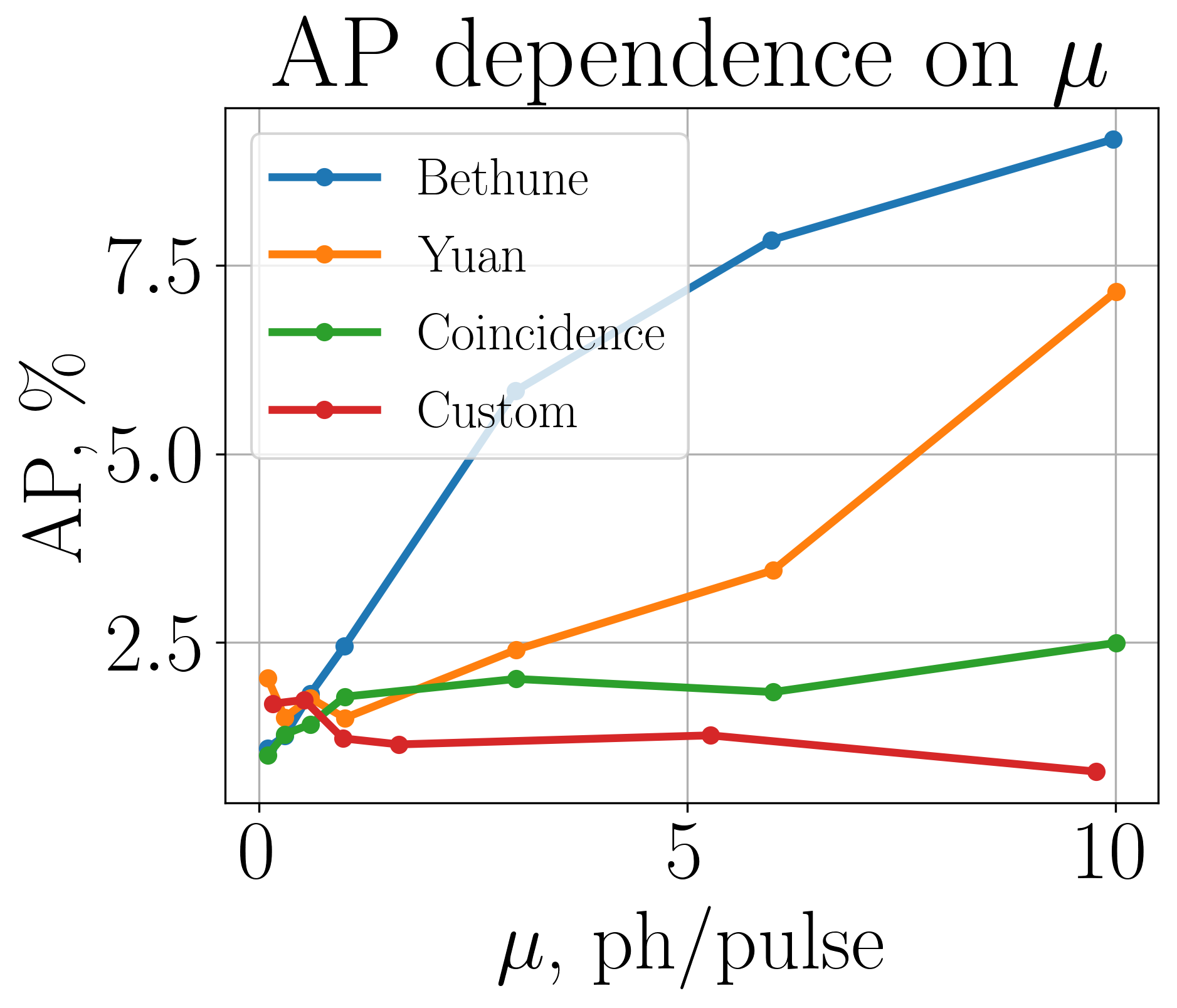}
   \caption{Afterpulse histogram for different methods: a) the Bethune method ($f_l = 156.25$ MHz); b) the Yuan method ($f_l = 6.25$ MHz); c) the coincidence method ($f_l = 6.25$ MHz); d) the comparison of the afterpulse probability for different methods to laser pulse energy. The green dashed line denotes the gate boundaries. Measurements were performed on custom SPD with the intensity of laser pulses $\mu = 0.1$ ph/pulse.}
   \label{fig:popular_methods}
\end{figure}

The Bethune, Yuan, and coincidence methods are pretty similar; however, they differ in histogram collections and postprocessing approaches. We need to collect two histograms in each method: with and without light illumination. In figure \ref{fig:popular_methods} we collect the histograms for each of the methods.

In the Bethune method, we need to set up the frequency of laser pulses as half of the gating pulses. For our SPD gating frequency  $f_g = 312.5$ MHz, we set up laser pulse frequency to $f_l = 156.25$ MHz. For more details about our custom SPDs and measuring stand, see the section \ref{sec:stand}. We need to collect histograms of detector triggers dependent on time with a resolution, preferably more than ten bins per gate, to make different gates distinguishable on histograms (as in the Yuan method). Afterpulsing obtaining with postprocessing of such histograms (see figure \ref{fig:popular_methods} a):

\begin{equation}
	P_{ap} = \frac{R_{ni} - R_{dark}}{R_{de}},
\end{equation}

where $P_{ap}$ is the afterpulse probability, and with illumination: $R_{ni}$ is count rate in non-illuminated gate $R_{de}$ is count rate in two consecutive gates; without illumination: $R_{dark}$ is count rate in gate.

In the Yuan method, we can set laser pulses frequency multiple of gate frequency, for example, one-fiftieth: $f_l = 6.25$ MHz. We collect the similar histogram as in the Bethune method (see figure \ref{fig:popular_methods} b), and post-process it:

\begin{equation}
	P_{ap} = \frac{R_{ni} - R_{dark}}{R^c_{de} - R_{ni}} \cdot \frac{f_g}{f_l},
\end{equation}

where $R_{ni}$ is the specific bin after the illuminated gate,  $R^c_{de}$ are the coincidence triggers of SPD and laser pulse arriving, i.e., count rate in the illuminated gate.

The main disadvantage of this method is that we are looking at afterpulsing only by one specific gate to calculate $R_{ni}$. However, afterpulsing has an exponential distribution over the gate number, and if we get another gate with own $R_{ni}$, then we obtain another result. 

In the coincidence method, we can solve the issue of the Yuan method. For measuring, we get the same laser pulse repetition rate: $f_l = 6.25$ MHz, and collect the histogram over the sweep, equal to one laser pulse period. We can use less resolution in this method because gates distinguishability is unimportant. We can process histogram (see figure \ref{fig:popular_methods} c) as follows:

\begin{equation}
	P_{ap} = \frac{R_{de} - R^c_{de} - (1 - f_l / f_g) R_{dark}}{R^c_{de}},
\end{equation}

where $R_{de}$ is the total count rate per one laser pulse period.

As we can see, all of these models describe the afterpulsing to varying degrees of accuracy. The afterpulsing is an internal SPD parameter, and it should not depend on external factors, like laser pulse repetition rate or average energy per laser pulse. Nevertheless, the afterpulsing is a non-markovian process \cite{wang2016non}, and its probability, obtained with standard methods, will strongly depend on the count rate (see figure \ref{fig:popular_methods} d). It means that each of described methods does not allow us to obtain actual internal afterpulse probability.

So these afterpulse models, despite being widely used for QKD systems, have considered disadvantages. However, our priority task is to determine the valid internal afterpulsing parameter, which response to a single triggering. We present the measurement technique and a probabilistic model for estimating the afterpulse based on the recursive nature of this effect. It will allow us to obtain valid internal afterpulsing parameters. We can model our device at any external parameters if we know valid internal parameters, like laser pulse frequency or average laser pulse energy. In the next section, we introduce our model.

\section{Custom afterpulse models}\label{sec:models}

\noindent Afterpulse is a detection event, that follows a previous detection event, is correlated to a previous detection event, and is not due to photon incident at detectors input \cite{wang2019afterpulsing}. Afterpulse is one of the SPD noise components, like dark counts. Dark counts don't correlate with triggering history and have several causes in the internal SPAD structure: charge band-to-band tunneling, trap-assisted tunneling, thermal generation and other less significant mechanisms \cite{tosi2014low}. For dark count rate we will use acronym $DCR$. 

Afterpulse has the following mechanism: after a detection event, the avalanche quickly quenches, and electrical current through device structure lowers too. After total avalanche quenching, there are a lot of trapped charges in the device structure, that relax with a time. If SPAD bias recovers, than there is high probability to trigger the avalanche due to a detrapped charge. If we forcibly hold the SPAD at off state (with low bias), than after some time, all charges will be detrapped, and new trigger event will not correlate with previous. However, usage of such regime for InGaAs/InP SPAD is impractical, due to high relaxation time. In this case, we have a compromise: if we want low afterpulse, we need to increase hold-off time, but our limiting count rate will lower. 

The afterpulse click can trigger the next afterpulse click (second-order afterpulses). This effect is negligible for low afterpulse probabilities ($p_{ap} < 5 \% $) but should be carefully accounted for high. In this section, we present two different models -- "Simple" and "Complex". The advantage of the simple model is its simplicity and possibility of easy $p_{ap}$ calculation from the statistics. The advantage of the complex model is that it is an accurate statistical description of afterpulse processes in the diode.

Here we assume, that probability of the afterpulse caused an earlier afterpulse is the same as the probability of the afterpulse caused by a laser pulse. We made this assumption in accordance with physical nature of afterpulsing effect. We can’t separate clicks, caused by afterpulse, or by laser pulse, because both avalanche processes start with one hole (in InGaAs/InP SPAD) in multiplication region. And there is no additional trapped charge in the heterostructure after the avalanche process, in one of the cases, because both avalanches are on average the same. But if our laser pulses consist of hundreds of photons, most likely we will be able to separate these avalanches.  In our model we consider only low-energy laser pulses ($\approx 0.1 - 10$ photons per pulse), and  with high-energy pulses our model is untenable.

The simple model's main idea is the afterpulse click probability $P^s_{ap}$ can be derived from photon-, thermal-, tunneling- induced click probabilities $P_0$ by the parameter $p^s_{ap}$. It means, that $P_0$ consists of light-induced clicks and dark count clicks.

\begin{equation}
	P^s_{ap} = P_0 p^s_{ap}.
\end{equation}

\begin{figure}[ht]\centering
	\includegraphics[width=0.47\linewidth]{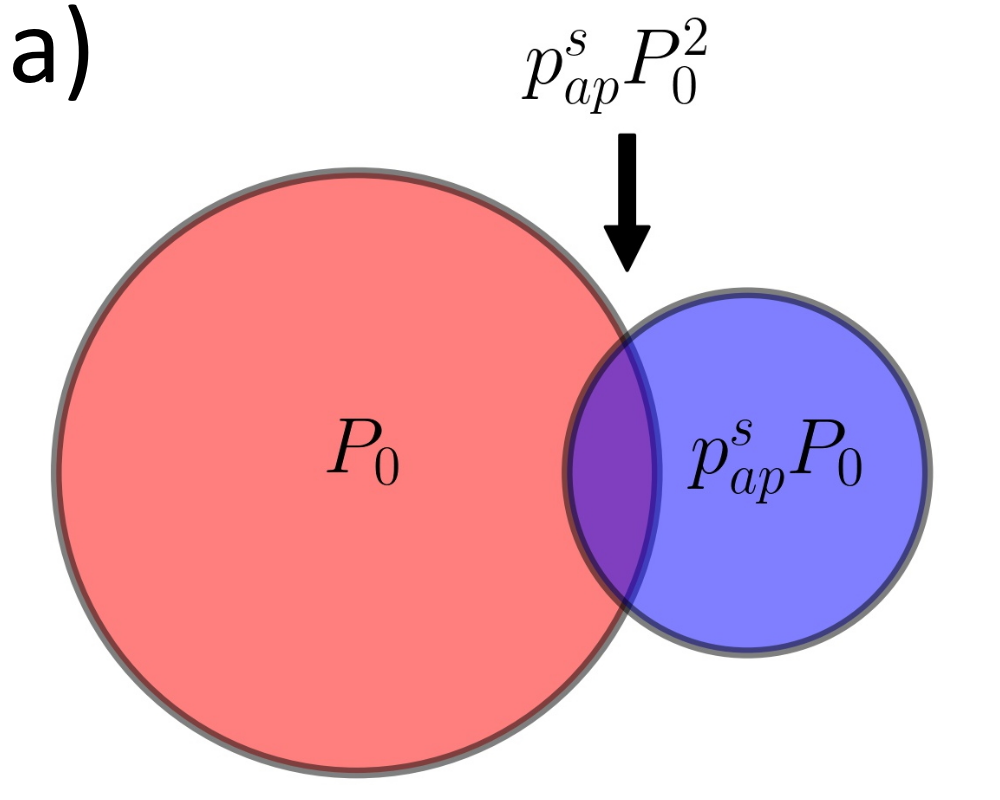} \hspace{3mm}
	\includegraphics[width=0.47\linewidth]{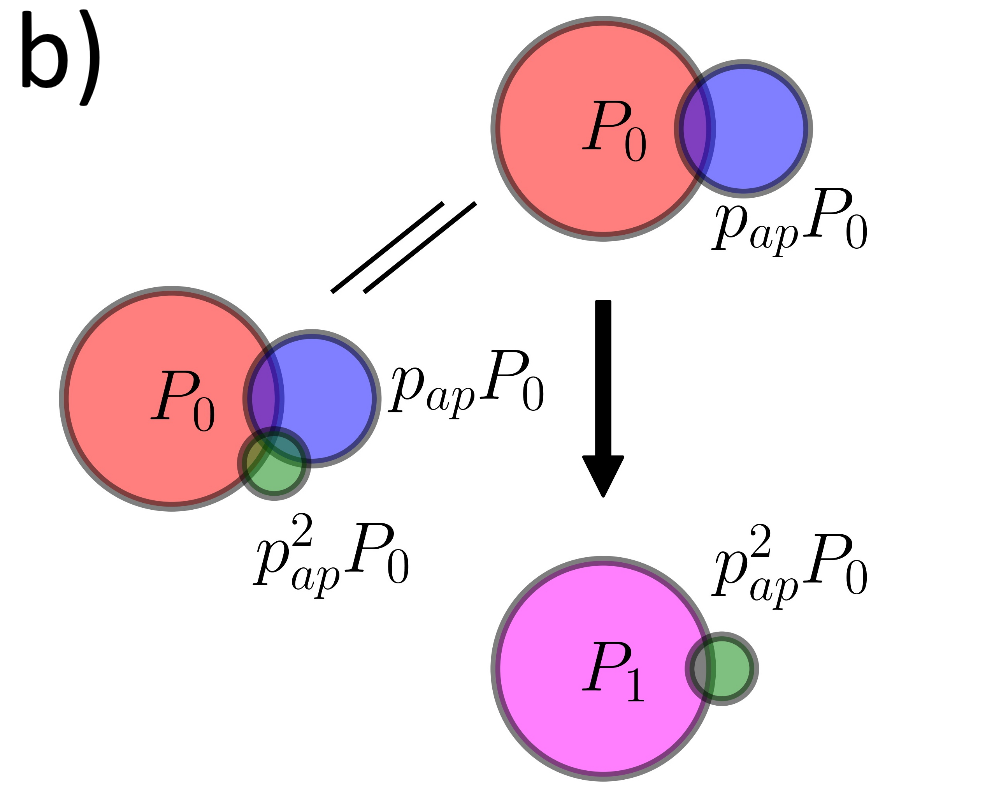}
   \caption{Schematic diagram of afterpulse probability models: a) simple model; b) complex model. $P_0$ is a probability of clicking in a system without afterpulse, $p_{ap}$ -- is the afterpulse probability.}
   \label{fig:diag1}
\end{figure}

In this view, the $P^s_{ap}$ included the second, third, etc. order afterpulses, and its assessment is included in the parameter $p^s_{ap}$. This approach is right for the low afterpulse probabilities because the high-order afterpulses are unlikely. The main drawback is that with varying the $P_0$ by, for example, an increase in the number of photons per pulse, the $p^s_{ap}$ assessment will differ too \cite{fan2020optimizing}. We can find the total probability of click $P$ from the schematic diagram \ref{fig:diag1} a).

The total click  probability $P$ can be derived as:

\begin{equation}\label{eq:simp}
	\text{Simple: } P = P_0 + P^s_{ap} - P_0 P^s_{ap} = P_0 (1 + p^s_{ap} - P_0 p^s_{ap})
\end{equation}

In the complex model, we take into account the recursive behavior of the afterpulse. In this case, we can show the next schematic diagram, presented in figure \ref{fig:diag1} b).

The probability that the clicks $P_0$ will generate an afterpulse is equal to $P_0 p_{ap}$. The probability that afterpulses $P_0 p_{ap}$ generate  new afterpulses is $P_0 p^2_{ap}$, and so on. The calculation of the overall afterpulse influence on the total probability of click can be performed by a series of consecutive convolutions of probabilities ($P_1$, $P_2$, etc.). We can write the next equations:

\begin{equation}
	\begin{split}
		&P_1 = P_0 + P_0 p_{ap} - P_0 P_0 p_{ap} = P_0 (1 - P_0 p_{ap}) + P_0 p_{ap}, \\
		&P_2 = P_1 + P_0 p^2_{ap} - P_1 P_0 p^2_{ap} = P_1 (1 - P_0 p^2_{ap}) + P_0 p^2_{ap}, \\
		&\vdots \\
		&P_{i+1} = ... = P_i (1 - P_0 p^{i+1}_{ap} ) + P_0 p^{i+1}_{ap}. \\
	\end{split}
\end{equation}

This recurrent equation can be rewritten as the decomposition relate parameter $P_0$. To do this, we consider the probabilities of afterpulse events as $\gamma_i$, where $P(\gamma_i) = P_0 p^i_{ap}$, and these events are joint and independent, which means that $P(\gamma_i \cap \gamma_j) = P(\gamma_i) P(\gamma_j) = P^2_0 p^{i + j}_{ap}$.

\begin{equation}
	\begin{split}
		P_1 &= P(\gamma_0 \cup  \gamma_1) = P(\gamma_0) + P(\gamma_1) - P(\gamma_0 \cap \gamma_1) = \\
		&= P_0 (1 + p_{ap}) - P^2_0 p_{ap},\\
		P_2 &= P(\gamma_0 \cup \gamma_1 \cup \gamma_2) = P(\gamma_0) + P(\gamma_1) + P(\gamma_2) - \\
		&- P(\gamma_0 \cap \gamma_1) - P(\gamma_0 \cap \gamma_2) - P(\gamma_1 \cap \gamma_2) + \\
		&+ P(\gamma_0 \cap \gamma_1 \cap \gamma_2) = P_0 (1 + p_{ap} + p^2_{ap}) - \\
		&- P^2_0 (p_{ap} + p^2_{ap} + p^3_{ap}) + P^3_0 p^3_{ap},\\
		\vdots \\
		P_n &= P(\gamma_0 \cup \gamma_1 \cup \gamma_2 \cup \hdots) = P_0 \sum^{n}_{i = 0} p^i_{ap} - \\
		&- P^2_0 \sum^n_{i,j = 0; j > i} p^{i + j}_{ap} + P^3_0 \sum^n_{i, j, k = 0; k > j > i} p^{i+j+k}_{ap} + \hdots
	\end{split}
\end{equation}

In this recursive equation, $P_{\infty}$ is the actual probability of the click, which includes  all orders of afterpulses. For ease of use in analytical models, we can take the first and second terms in the appropriate order of $P_0$. After that, we will analyze the bounds of applicability of these two decomposition models.

We can calculate the sum of the series of first and second order $P_0$ as follows:

\begin{equation}
	\begin{split}
		&\lim_{n \to \infty} \sum^n_{i = 0} p^i_{ap} = \frac{1}{1 - p_{ap}},\\
		&\lim_{n \to \infty} \sum^n_{i, j = 0; j > i} p^{i + j}_{ap} = \frac{p_{ap}}{(1 - p_{ap})^2 (1 + p_{ap})}.
	\end{split}
\end{equation}

So, we can derive the first and second-order afterpulse accounting models as:

\begin{equation}\label{eq:2}
	\begin{split}
		\text{1st order}:& \ P = P_0 \frac{1}{1 - p_{ap}}, \\
		\text{2nd order}:& \ P = P_0 \frac{1}{1 - p_{ap}} - P^2_0 \frac{p_{ap}}{(1 - p_{ap})^2 (1 + p_{ap})}. \\
	\end{split}
\end{equation}

\begin{figure}[ht]\centering
	\includegraphics[width=0.47\linewidth]{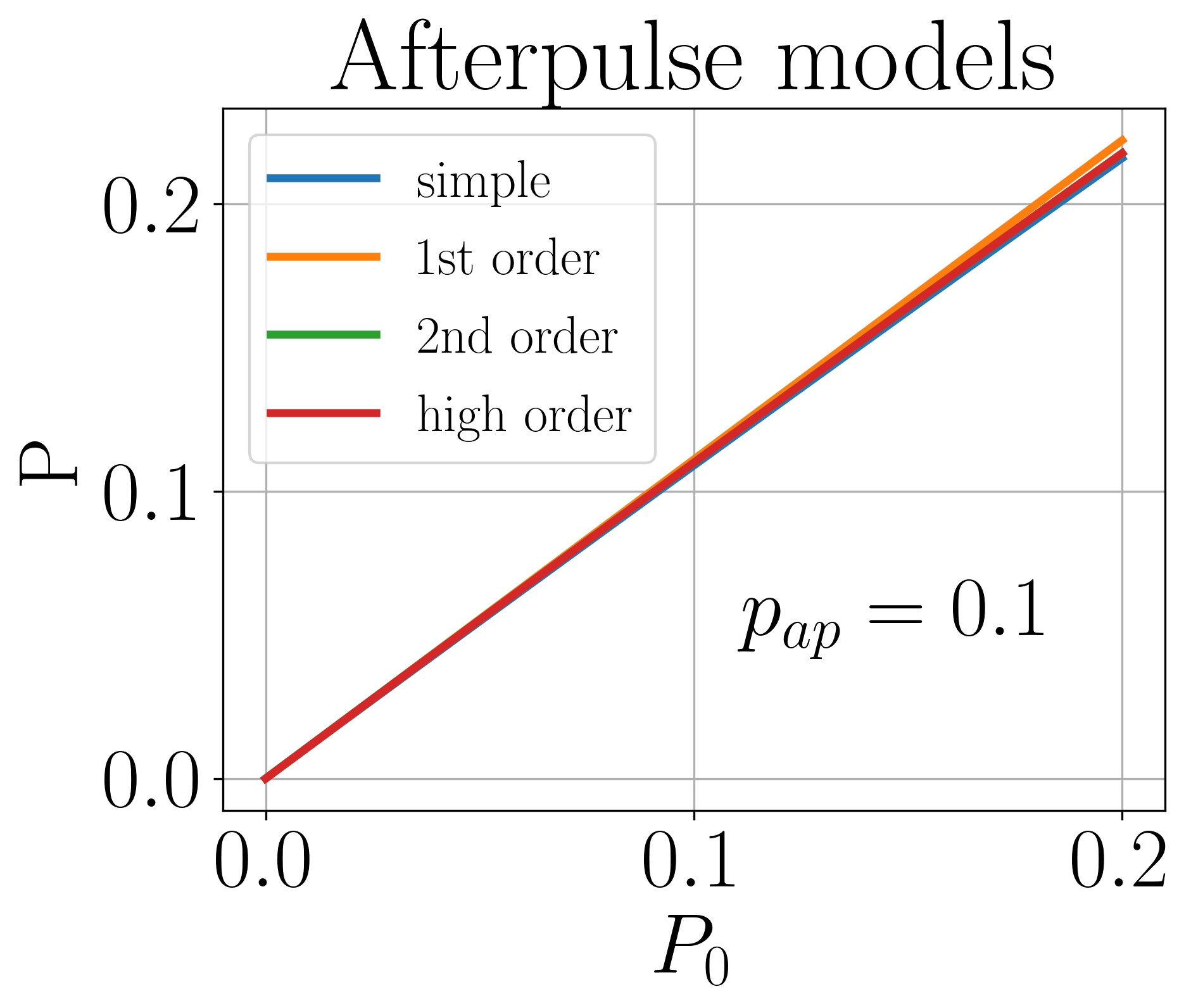} \hspace{3mm}
	\includegraphics[width=0.47\linewidth]{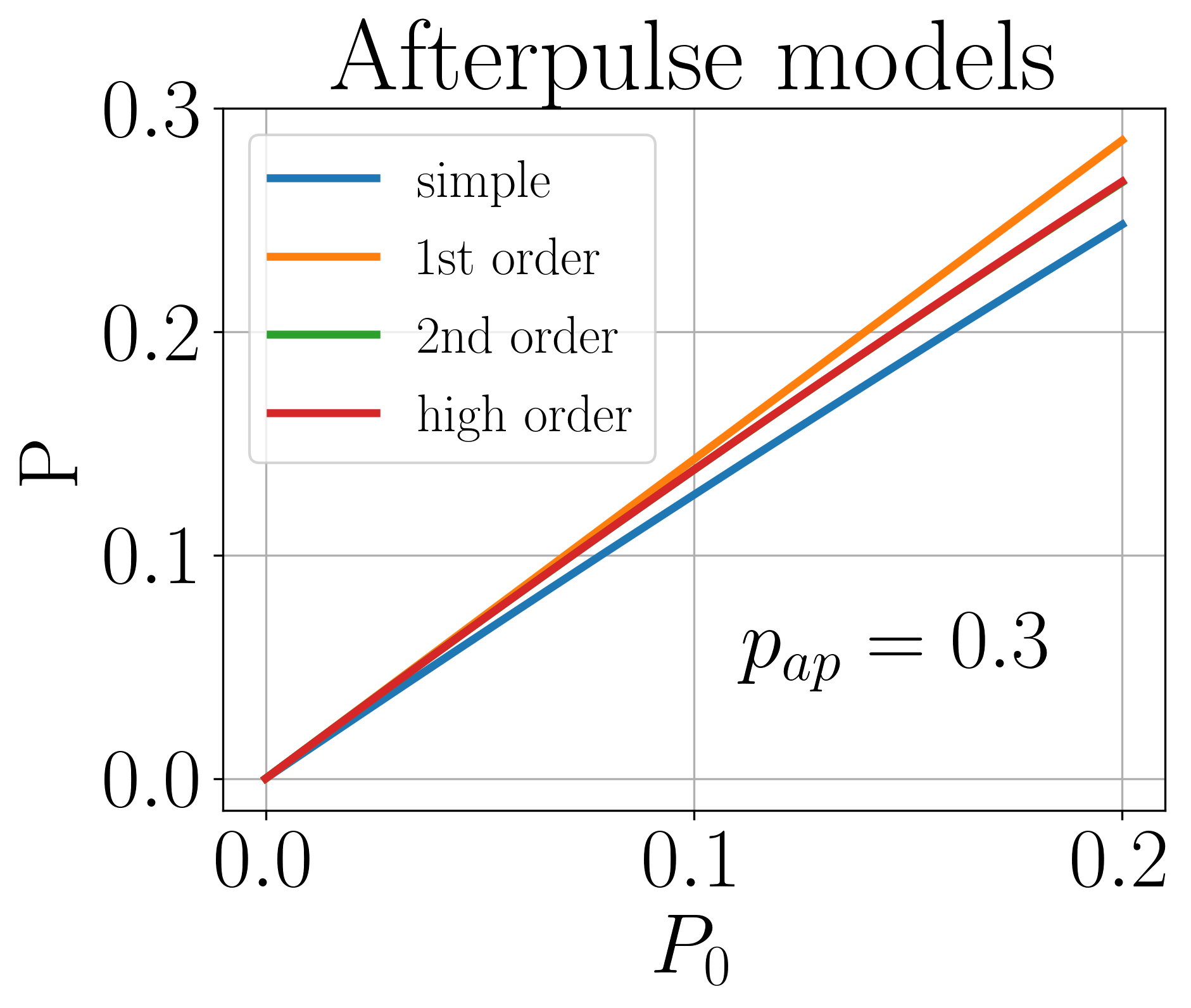}
   \caption{Comparison of the simple, 1st, 2nd and high order models for determining the click probability $P$, assuming that a) $p_{ap} = 0.1$, b) $p_{ap} = 0.3$.}
   \label{fig:pap_calc}
\end{figure}

In figure \ref{fig:pap_calc}, we compare the detection probability $P$, calculating according to simple, first, second, and high order models. We assume that  high-order model (the decomposition of $P$ with 20th order of $P_0$) is the benchmark, and we should compare the other models to it. We can see that for low afterpulse probability $p_{ap} = 0.1$  all models give  good convergence with the high order model, and the low deviations begin with increasing the $P_0$. The first model gives the largest error. However, for the higher values of $p_{ap}$, we can see that simple and first-order have large deviations. Only a second-order model should be used for accurate estimation of total click probability $P$.

\section{Influence of the active reset on the counting statistics}\label{sec:stand}
 
\begin{figure}[ht]\centering
	\includegraphics[width=0.8\linewidth]{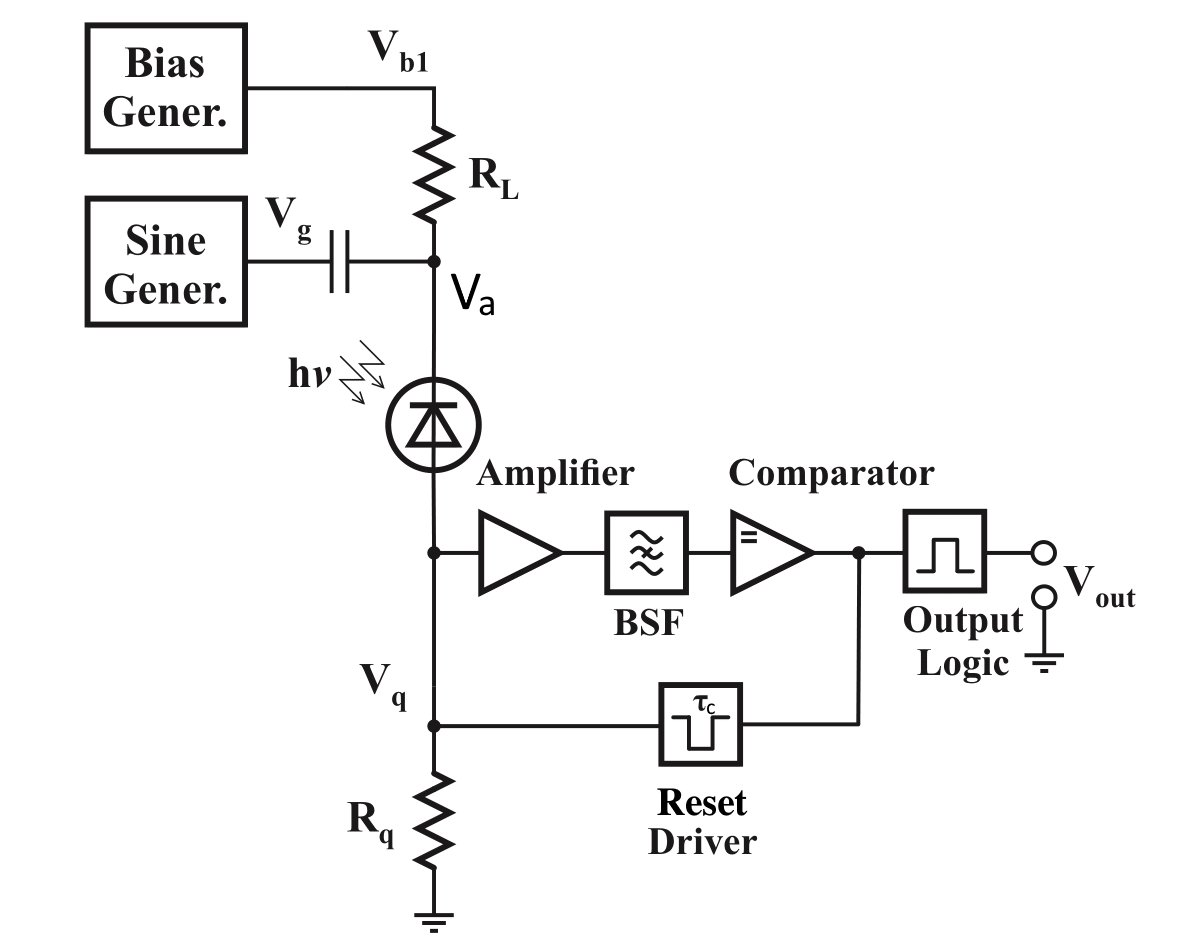}
   \caption{Functional circuit of SPD. In the circuit: Bias Gener. -- DC bias generator; Sine Gener. --  AC sinusoidal bias voltage (gate) generator with a frequency of $312.5$ MHz; BSF -- the filters to eliminate the influence of gates on the processing of avalanche signals. }
   \label{fig:el_sch}
\end{figure}

\noindent The principal SPD scheme is presented in figure  \ref{fig:el_sch}. Here, bias generator applies a DC bias, and sine generator applies the sinusoidal bias to SPAD. The sinus frequency is $312.5$ MHz. Such frequency used due to SPD was developed to use in our QKD device, that works with same laser pulses repetition rate.    Arriving photon can trigger the self-sustaining avalanche process, that leads to high current through SPAD and consequent voltage drop can be registered by comparator and classified as click.   The quenching process is multistage. The first quenching realized by gated signal -- when bias drops below breakdown voltage. After that resistor $R_q$ ($\approx 50 \ \Omega$)  passively quenches the avalanche, that  for some time is still growing in the SPAD linear regime.  This resistor takes up part of the voltage across the diode, due to sufficiently increased current.

But if we use only the passive quenching, after about hundreds of nanoseconds, the new avalanche processes can occur. It's due to relaxation of DC bias voltage, determined by resistor $R_l \approx 47 \ k\Omega$, and circuitry and design features of the implementation, introducing parasitic reactances.  Moreover, there are a lot of high amplitude relaxation modes after avalanche signal passes through amplifiers and filters block, that can lead to comparator re-triggering (it continues about $100$ ns). To prevent this we add a latching to comparator. During latching time $\tau_l$ we can`t observe SPD triggers, and we can consider it as dead time (we named it "Latched time" (LT)). In figure \ref{fig:el_sch} this regime corresponds to scheme without reset driver, and therefore it's relatively simple for circuit realisation. The main disadvantage is that new avalanches can grow and passively quench during latching time, that sufficiently increases trapped charges, and therefore the afterpulse probability.

To avoid problems with many avalanche triggers during the dead time, after the registration event, we need to use active hold-off time -- this is the additional stage in quenching process.  We need to note, that this scheme will still be the passive quenching, because first quenching is due to passive elements. Active hold-off enabling delay is about $10$ ns after comparator triggering.  In figure \ref{fig:el_sch} quenching driver applies the electrical pulse with width $\tau_c$ (schematic dead time), that significantly drops the voltage on the SPAD. When SPAD is biased under breakdown, it can't be triggered by single photons, and therefore charge traps continue to relax. The reset to the normal operating regime of the SPAD is due to change of the applying voltage and can be seen as active reset.  This dead time realization we named "Latched time + active reset" (LT + AR)  \cite{liu2020ultra, liu2021exploiting}.

Also, we can define  statistical dead time $\tau_s$ as the time range between some trigger and the next first possible trigger. This time interval depends on $\tau_l$, $\tau_c$, and active reset pulse form. This pulse should be square in the ideal case, but the leading and trailing edges are distorted due to electronics influence.

Trail edge causes high amplitude oscillations, which are difficult to learn due to their high-frequency components, making detector performance difficult to predict. To exclude possible counting of these triggers, we need to make latched time $\tau_l$ more than schematic dead time $\tau_c$ on the value, approximately equal to the time of these transients $\tau_{er}$. This recovery time can be in the range $200 < \tau_{er} < 3 \ \mu s$, which depends on the circuit realization, but low values are preferable.

\begin{figure}[ht]\centering
	\includegraphics[width=0.47\linewidth]{hardpic1.png} \hspace{3mm}
	\includegraphics[width=0.47\linewidth]{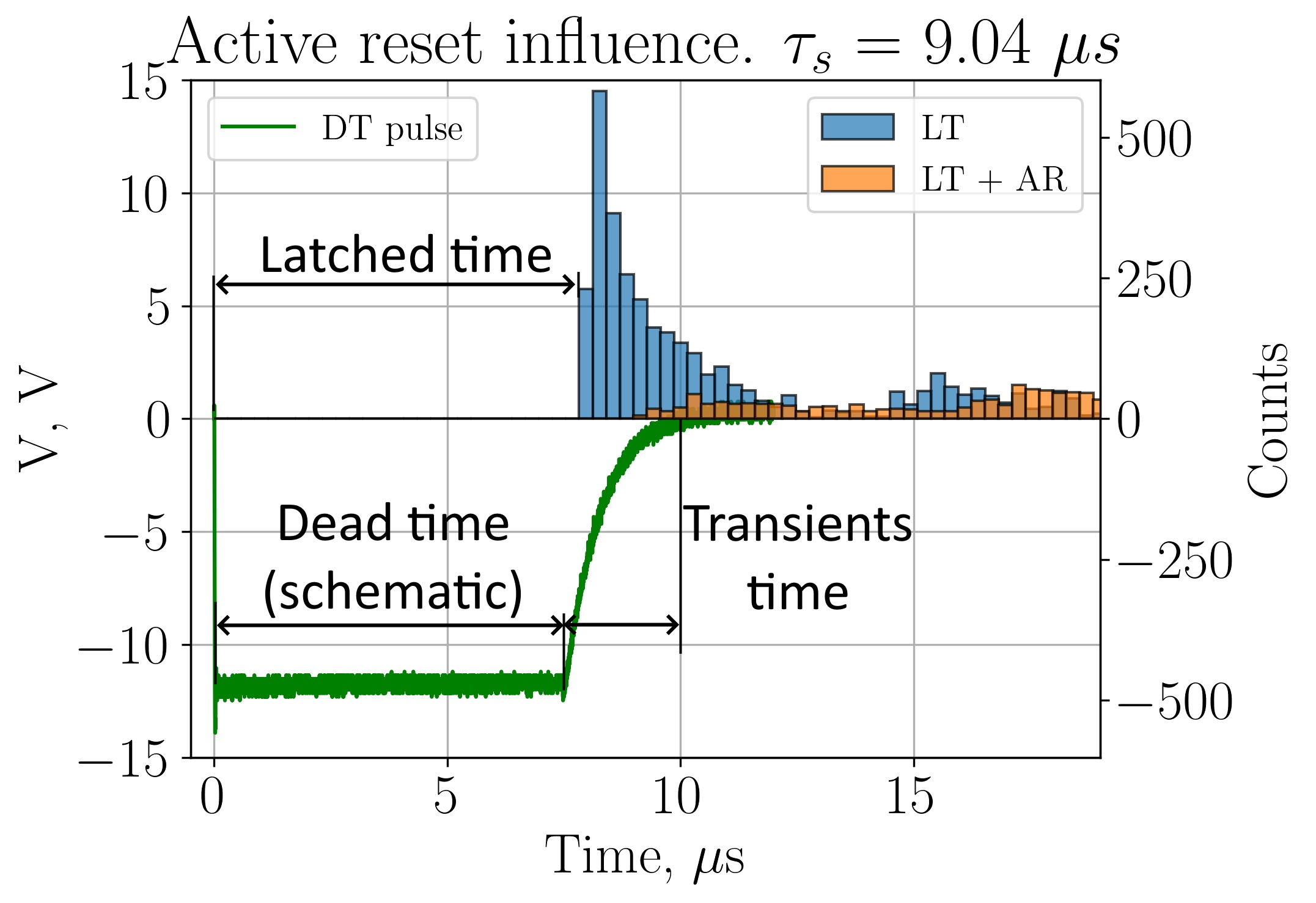}
   \caption{Active reset influence on the counting statistics: a) $\tau_s = 5.35 \ \mu s$; b) $\tau_s = 9.04 \ \mu s$. Two statistics were normalized by zero bin click count $C_0 - C_{dcr}$.}
   \label{fig:hardpic1}
\end{figure}

There are different configurations that depend on these time intervals:

\begin{itemize}
	\item $\tau_l > \tau_c + \tau_{er}$: we present this case in figure \ref{fig:hardpic1} a). This figure presents the active reset time and counting statistics for LT and LT + AR schemes (as for the b). Here, latched time fully determines statistical dead time $\tau_s$: $\tau_s = \tau_l$. Here, $\tau_{c} \approx 3.65 \ \mu s$ and $\tau_{er} \approx 0.8 \ \mu s$. The difference $\tau_l - \tau_c - \tau_{er} \approx 0.9 \ \mu s$. It means that approximately $0.9 \ \mu s$ in the DT + AR scheme did not detect triggers can occur.
	
 	\item $\tau_l < \tau_c + \tau_{er}$: we present this case in figure \ref{fig:hardpic1} b). Here, $\tau_c \approx 7.5 \ \mu s$, $\tau_{er} \approx 2.5 \ \mu s$, $\tau_l \approx 7.8 \ \mu s$. We can see that the first triggers in the LT + AR scheme occur at $9 \ \mu s$, which is lower than $\tau_c + \tau_{er} = 10 \ \mu s$. The statistics are due to the SPD's low detection probability when active reset pulse applied to SPAD and its value $< 0$ because it lowers the bias voltage. The deviation of the initial section of statistics from the mentioned afterpulse laws is determined by the photon detection efficiency ($PDE$) dependence on the SPAD bias voltage.
\end{itemize}

The case, in which using the LT + AR scheme and satisfied $\tau_l \approx \tau_c + \tau_{er}$, where $\tau_{er} \approx 400$ ns is preferable. There are no triggers before the latched time to increase the trap's charge and, consequently, the afterpulse probability. However, setting such a value is a rather non-trivial schematic and technical task.

\section{Afterpulse measurement approach}\label{sec:measurements}

\begin{figure}[ht]\centering
	\includegraphics[width=1\linewidth]{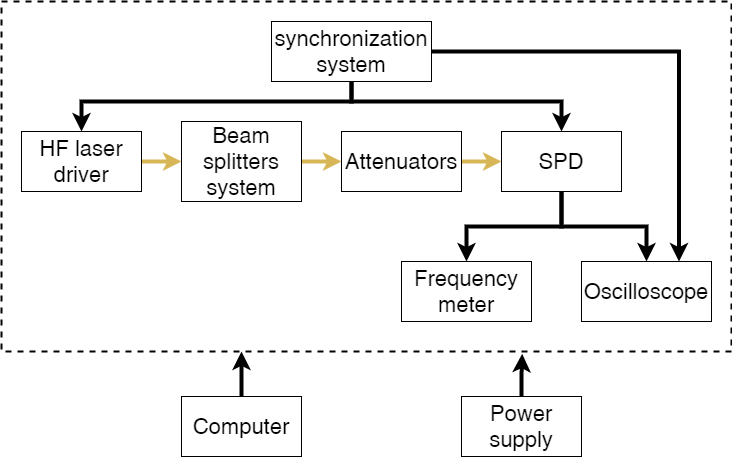} 
   \caption{Functional scheme of SPD’s measurement setup: frequency meter – Keysight 53230A, oscilloscope – Lecroy WaveMaster 830Zi-B-R.  }
   \label{fig:stand_scheme}
\end{figure}

\begin{figure}[ht]\centering
	\includegraphics[width=0.47\linewidth]{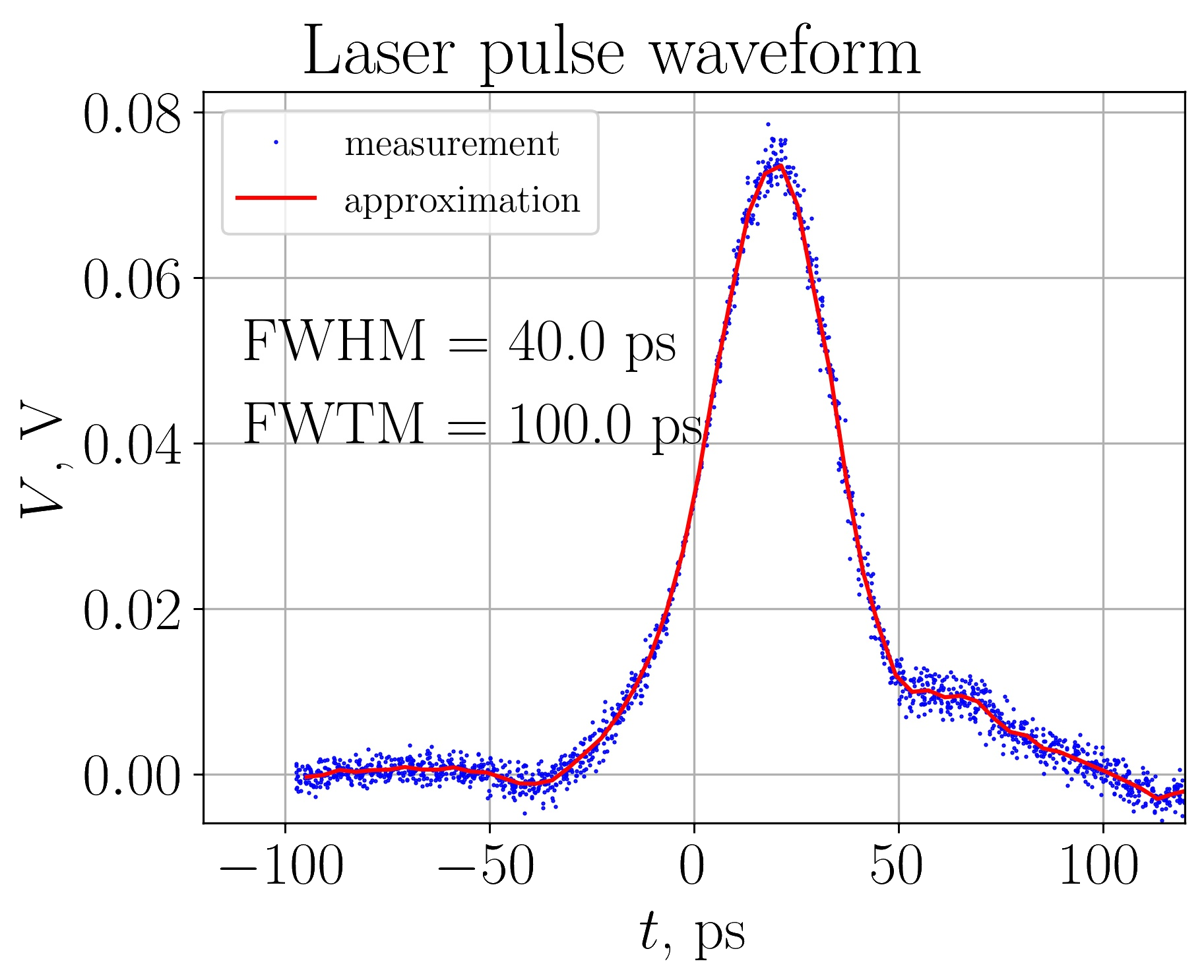} 
	\includegraphics[width=0.47\linewidth]{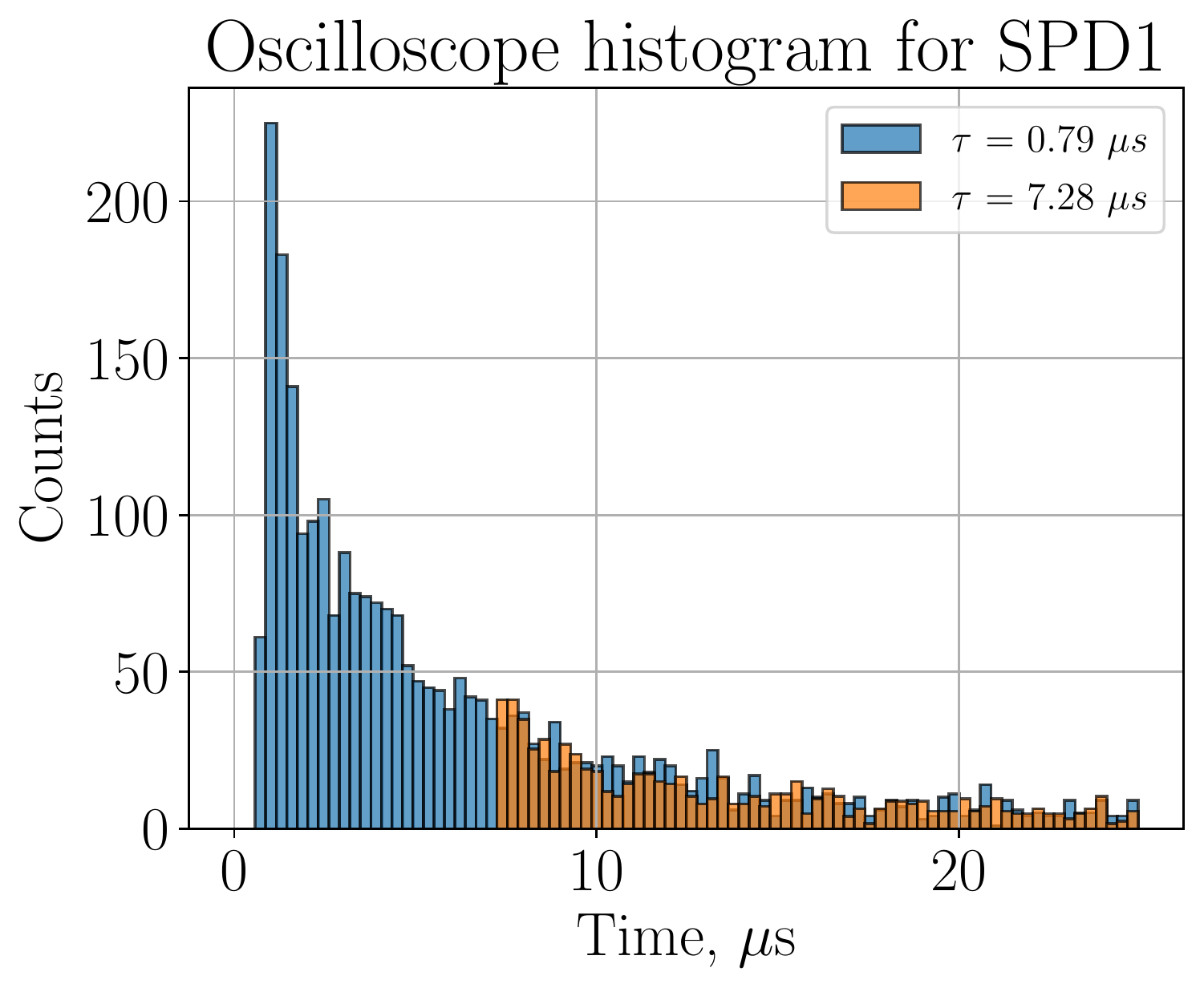}
   \caption{a) experimental laser pulse waveform; b) experimental histogram of the triggers for SPD1, measured for two different dead times: $\tau = 0.21 \ \mu s$ and $\tau = 7.3 \ \mu s$.}
   \label{fig:laser}
\end{figure}

\noindent For our measurements we use experimental stand, which principal scheme is presented at figure \ref{fig:stand_scheme}. Black arrows denote electrical connection, yellow denote optical connection. Experimental stand consists of synchronization system, that outputs synchronization signals to high frequency (HF) laser driver, single photon detector and oscilloscope. Due to SPD is gated, we need to synchronize it with laser pulses.  HF laser driver controls temperature-stabilized laser, and its output  optical laser signal with wavelength $\lambda \approx 1550$ nm, width about $100$ ps and FWHM about $40$ ps, presented at figure \ref{fig:laser} a). In our measurements we used laser pulse repetition rate $10$ kHz.
 
Laser pulses arrive to beam splitters system, and one of its outputs  connected to attenuators. Other outputs are used to control laser pulse waveform. Attenuators block consists of two devices: first one has output power control, and maintains output power by changing it's attenuation; and the second holds a stable attenuation. 

After the attenuators, each laser pulse has mean power about 1 photon. Such signal arrives to SPD, and causes it to be triggered. SPD triggers output electrical signal to frequency meter and oscilloscope. With frequency meter we can evaluate the $DCR$ and $PDE$ parameters, and on oscilloscope time resolution and afterpulse histograms. 

By the computer we can control each block of this scheme, instead of beam splitters system, that is passive. Power supply powers each active element.

\begin{figure}[ht]\centering
	\includegraphics[width=1\linewidth]{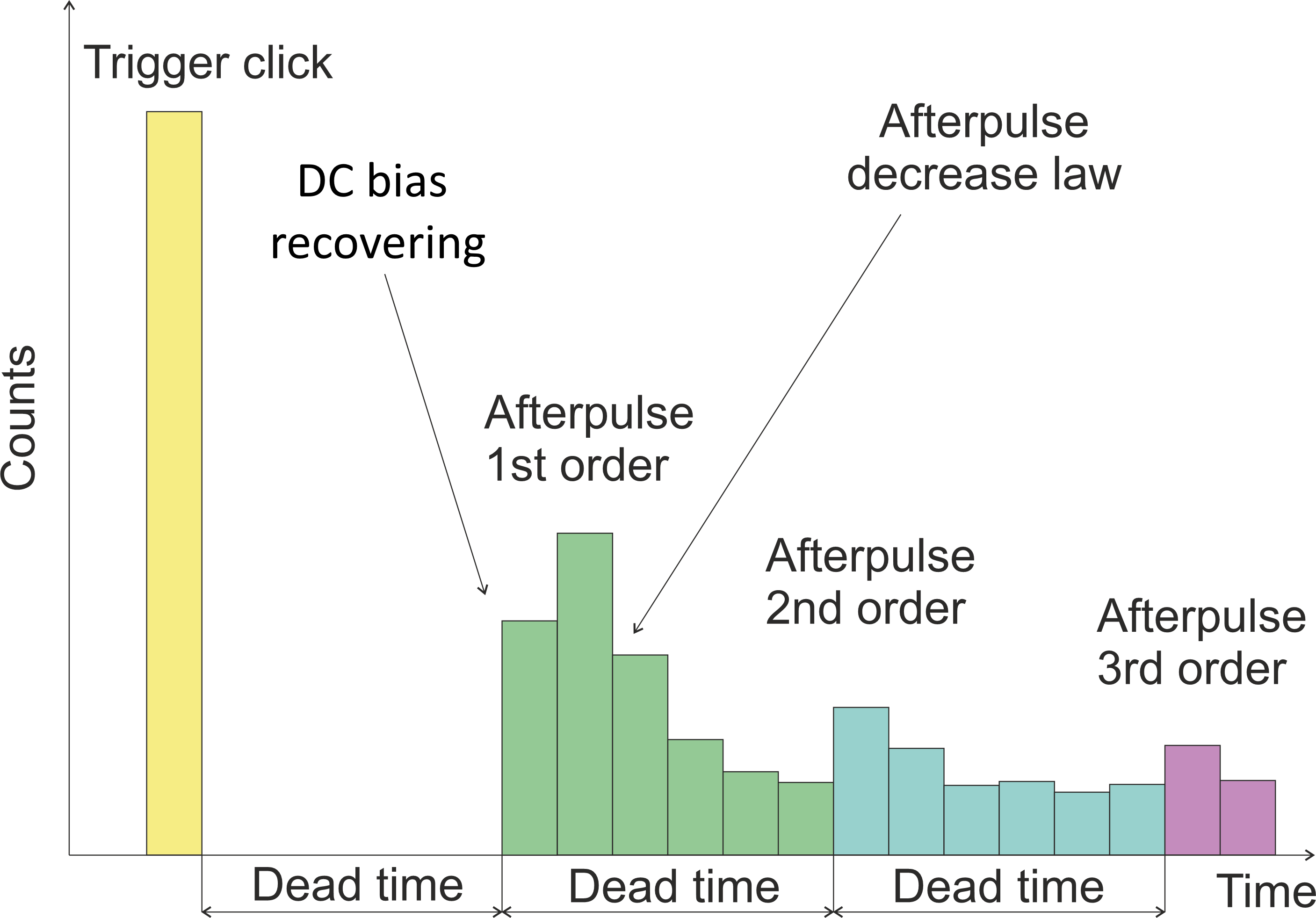}
   \caption{Schematic collected histogram.}
   \label{fig:hist_1}
\end{figure}

In our experimental approach, afterpulse measurements can be done by processing the triggers histogram, presented in figure \ref{fig:hist_1}. In the experimental setup, We set the oscilloscope sweep at $25 \ \mu s$. The detector is most likely to be triggered by laser pulses, and on this detection event oscilloscope will trigger. It's due to the fact that laser pulses period is lower then oscilloscope sweep. But, SPD can be triggered by internal noise, like $DCR$ or afterpulsing, and we need to  take this into account. 

After the trigger click bin, we have empty bins during the dead time. In the detector with LT + AR scheme with  well-established latched time of the comparator, new clicks can occur when DC bias voltage fully recovered. With our setup parameters, these clicks most likely are the afterpulses or dark counts, and not the light-induced.  In real case, adjustment  of latching time is quite complicated, and clicks can occur at growing edge of DC bias. Probability of such clicks is lower, than for fully recovered bias, and we observe this effect on the oscilloscope histograms like reduced some first bins. This effect is not due to poor-quality processing of the histogram. We collect statistics with $10$ ns oscilloscope bin with, and after that perform it merging to obtain  suitable data representations.

The afterpulse histogram slope could be described by one of the well-researched laws: exponential \cite{korzh2015afterpulsing, humer2015simple}, power-law \cite{itzler2012power}, or hyberbolic sinc model \cite{horoshko2017afterpulsing, ziarkash2018comparative}. In these works dead time was established to a fixed value, and afterpulse properties were obtained from histogram analysis. Nevertheless, in practice, this is a non-trivial task to eliminate effect of DC bias recovering influence on click probability, and the first bin after dead time presents this effect. In the third and other dead time windows, previous afterpulse clicks' influence distorts the representation of afterpulse law.

At the  figure \ref{fig:laser} b) we present the experimental histogram without trigger click bin, that suitable for afterpulse model fitting.   First dead time is $\tau = 0.21 \ \mu s$, and the second $\tau = 7.3 \ \mu s$. The measurement time for these two histograms is equal, and we can see that for time $t > 8 \ \mu s$, the bins high slope are roughly the same, which speaks in favor of the exponential afterpulsing law.

At the end of the oscilloscope sweep (the range $t_{dcr} \approx [20, 25] \ \mu s$), the count of afterpulse clicks will be negligible, and we can consider that it is mainly due to the dark counts. There is low probability that photon clicks in this time window because of the last's low repetition rate. We will argue that all counts that differ from the dark counts have  afterpulse nature for the rough evaluation. Furthermore, the total afterpulse counts can be found as:

\begin{equation}
	\begin{split}
		&C_{dcr} = \frac{1}{N_{dcr}} \sum_{i \in t_{dcr}} C_i, \\
		&C_{ap} = \sum_{i \in [\tau, 25] \ \mu s} (C_i - C_{dcr}), 
	\end{split}
\end{equation}

where $C_{dcr}$ -- is the average dark counts accumulated per bin, $N_{dcr}$ -- is the bins count in time interval $t_{dcr}$,  $C_i$ -- is the count in $i$-th histogram bin,  $C_{ap}$ -- is the total counts, that we consider as afterpulse counts (include 1st, 2nd etc. orders). 

The $C_{ap}$ counts include   different orders of  afterpulses. We estimate the afterpulse probability as:

\begin{equation}
	p^{exp}_{ap} = \frac{C_{ap}}{C_{0}},
\end{equation}

where the $C_0$ -- is the counts in trigger zero bin of histogram. 

The sense of the estimated value can be described by figure \ref{fig:diag1} a): $p^{exp}_{ap} P_0 \approx P_{ap} - P_{ap} P_0 $. Moreover, we can derive the $P_{ap}$ value that should be universal for all models  $P^{exp}_{ap} = p^{exp}_{ap} P_0 = P_{ap}$ as:

\begin{equation}
	P_{ap} \approx \frac{p^{exp}_{ap} P_0}{1 - P_0}
\end{equation}

We can easily calculate the $p_{ap}$ for simple, single, and second-order afterpulse models with this assumption. We can relate the $p^{exp}_{ap}$ and $p^s_{ap}$ and $p_{ap}$ parameters:

\begin{equation}
	\begin{split}
		&P_{ap} = \frac{P_n - P_0}{1 - P_{0}}, \\
		&p^{exp}_{ap} P_0 = P_n - P_0, \\
		&P_n = P_0 (1 + p^{exp}_{ap}),
	\end{split}
\end{equation}

where we obtained the right-hand side for the first equation from the simple equation for $P_n$ ($P$ notation for simple model) probability: $P_n = P_0 + P_{ap} - P_0 P_{ap}$. The $P$ and $P_n$ can be derived from equations \ref{eq:simp} and \ref{eq:2} for simple, first and second-order models.

For the simple and first order models, we can get the simple relation for $p^s_{ap}$ and $p^{(1)}_{ap}$:

\begin{equation}
	\begin{split}
		&p^s_{ap} = \frac{p^{exp}_{ap}}{1 - P_0}, \\
		&p^{(1)}_{ap} = 1 - \frac{1}{1 + p^{exp}_{ap}}. 
	\end{split}
\end{equation}

Similarly, we can find the $p^{(2)}_{ap}$ for the second-order model, but this is a non-trivial task to do it analytically. For this reason, $p^{(2)}_{ap}$ can be found only numerically.

We can find the $p^{(1)}_{ap}$ value from experimental $p^{exp}_{ap}$, but to calculate the $p^s_{ap}$ and $p^{(2)}_{ap}$, we need to find the $P_0$ probability previously. To do this, we need to know the overall probability of click $P_n$ (or $P$ for simple model), which we can find from count rate $R$ and dead time (statistical) $\tau$: $P_n = R \tau$ \cite{koziy2021investigating}. After that, we need to paste it into equation \ref{eq:simp} or \ref{eq:2} (depends on the observed model) and calculate the $P_0$. For the simple model, we can derive the $p^s_{ap}$ value from $R$ analytically:

\begin{equation}\label{eq:paps}
	p^s_{ap} = \frac{p^{exp}_{ap} (1 + p^{exp}_{ap})}{1 + p^{exp}_{ap} - R \tau}.
\end{equation}

It is evident that with low $R \tau$ the equation \ref{eq:paps} takes the following form: $p^s_{ap} = p^{exp}_{ap}$. 

For  accurate SPD models, one should use the second-order model because $p^{(2)}_{ap}$ does not depend on the $P_0$, and one value of $p^{(2)}_{ap}$ can be used for a wide range of optical power per pulse $\mu$ (ph/pulse). The simple model is suitable only for  low values of $p^s_{ap}$, and when SPD operation is proposed only with fixed $\mu$. First-order model suits in the case of low $p^{(1)}_{ap}$ because, like for the 2nd order model, it can be used for wide (but lower than for 2nd model) range $\mu$, and simple form of equations allows you to use it in analytical models of SPD.

\section{Results} \label{sec:results}

\noindent We tested three custom sinusoidal gated SPDs based on InGaAs/InP SPADs named SPD1, SPD2, and SPD3. These SPADs (with gated frequency $\nu = 312.5$ MHz) were manufactured by Wooriro company and were taken from different batches. The SPAD №2 and №3 have butterfly housing and a built-in enclosure cooling system. We have shown its main parameters in table \ref{table:0}.

\begin{table}[h]
	\caption{Parameters of used SPADs.}
	\label{table:0}
	\begin{center}
		\begin{tabular}{|c|c|c|c|c|}
			\hline
			\multirow{2}*{SPD №} & \multirow{2}*{SPAD} & \multirow{2}*{$T$, K} & \multicolumn{2}{c|}{$DCR$, Hz ($\tau \approx 20 \ \mu s$)} \\
			\cline{4-5}
			{} & {} & {} & $PDE = 10 \ \%$ & $PDE = 20 \ \%$ \\
			\hline 
			1 & {PA19H262-0004} & {223} &  $\approx 250$ & $\approx 450$   \\
			\hline
			2 & {MF20C300-0001} & {233} & {$\approx 50$}  & {$\approx 100$}  \\
			\hline
			3 & {MF20D300-0001} & {233} & {$\approx 150$}  & {$\approx 300$}  \\
			\hline
		\end{tabular}
	\end{center}
\end{table}

We added the possibility of enabling or disabling reset driver (passive-active quenching and reset) to the circuit realization of SPD. Also, we added the manual setting of the circuit dead time $\tau_{c}$ and latching time $\tau_l$. In all presented below figures, it was  $\tau_l > \tau_c$. However, due to transients at the trailing edge during the $\tau_{er}$, we cannot accurately set the parameters to achieve the preferable condition: $\tau_l > \tau_c + \tau_{er}$. For this reason, some data was obtained for the first case and another for the second case of time interval configurations, presented above. It means that statistical dead time $\tau_s$ for LT and LT + AR schemes can differ for the similar setting of latched time.

In our experiments, laser pulses with $\text{FWHM} \approx 50$ ps and repetition rate $10$ kHz were attenuated to the average power per pulse $\mu \approx 1$ ph/pulse. We established two $PDE$ in our experiments: $10 \ \%$ and $20 \ \%$. The setting of the $PDE$ value was performed by changing the SPD's bias voltage with unchanged gate amplitude. The statistics was collected on an oscilloscope and processed to obtain $p^{exp}_{ap}$  value. After that, we found $p^s_{ap}$, $p^{(1)}_{ap}$ and $p^{(2)}_{ap}$, which correspond to simple, first, and second-order models.

\begin{figure}[h]\centering
	\includegraphics[width=0.45\linewidth]{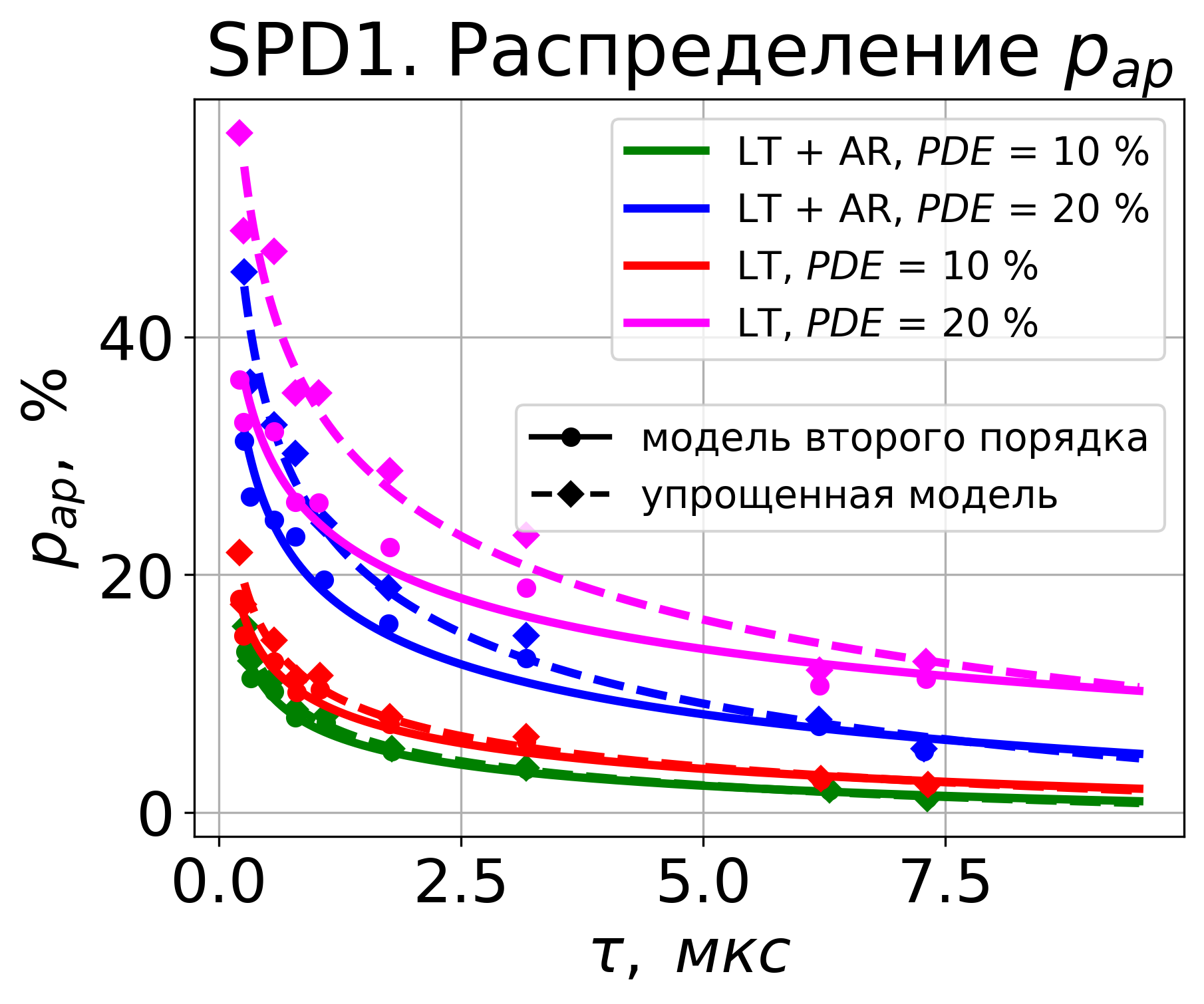} \hspace{3mm}
	\includegraphics[width=0.45\linewidth]{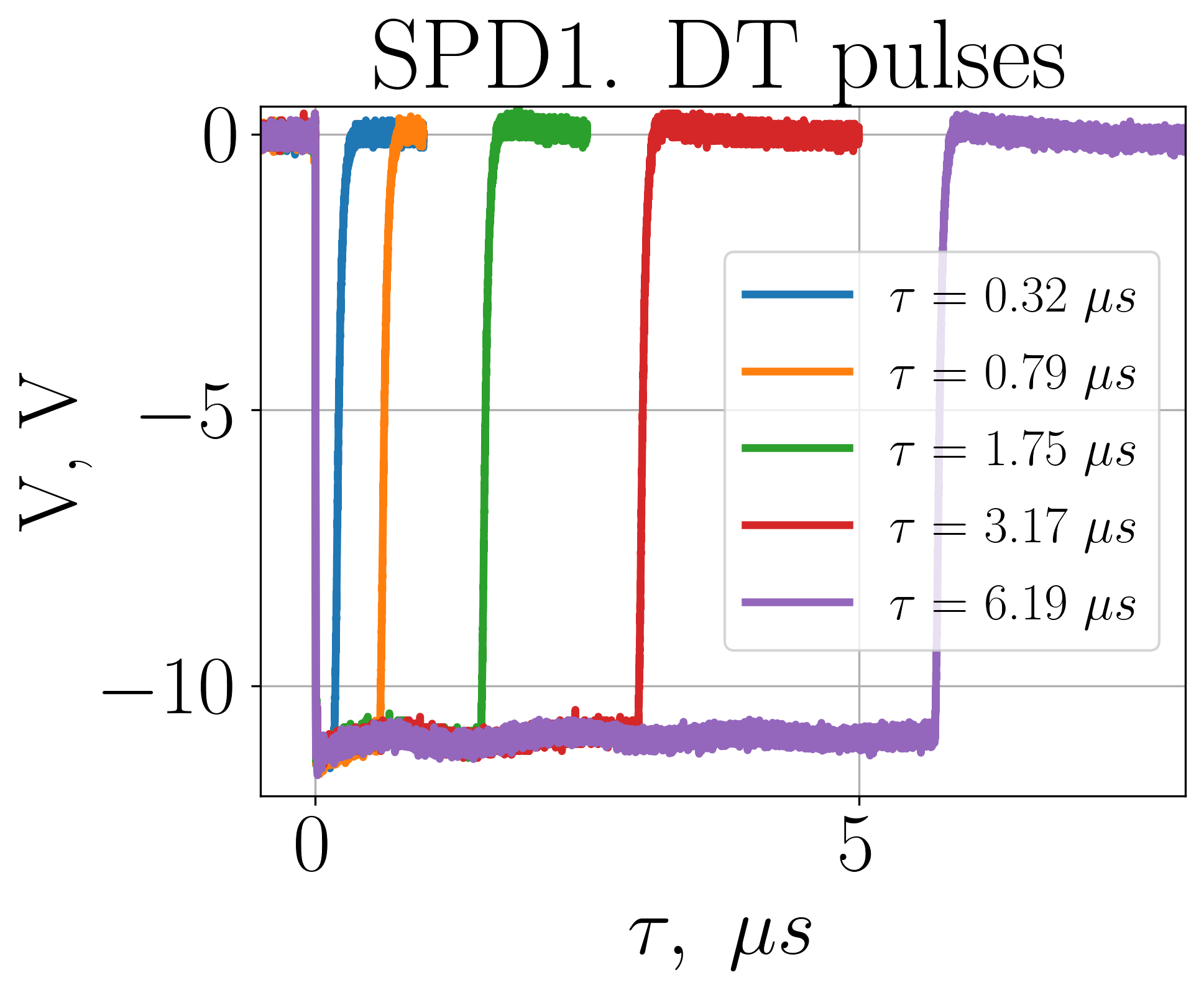}
   \caption{SPD1 data: a) Comparing $p_{ap}$ for LT and LT + AR schemes, obtained in according to simple and second-order models (diamonds and points correspondingly). Dashed lines represent the fitted curves for simple model, solid lines -- for the second-order model; b) Active reset pulse, applied to SPAD after the trigger. }
   \label{fig:apdt_spd1}
\end{figure}

In figure \ref{fig:apdt_spd1} a), we present the $p_{ap}$ value for LT and LT + AR schemes obtained for simple and second-order models (diamond and point markers correspondingly). We approximated data for SPD1 and SPD2 curves defined by the power-law equation. Data for SPD3 was approximated with an exponential equation:

\begin{equation}
	\begin{split}
		\text{power law: } &p^{pl}_{ap}(\tau) = A \tau^{B} + C, \\
		\text{exponential: } &p^{e}_{ap}(\tau) = A e^{-B \tau} + C,
	\end{split}
\end{equation}

where the $A$, $B$ and $C$ parameters have been fitted.

The common used law for approximate the $p_{ap}$ dependence on the dead time for fixed $PDE$ is the exponential, as shown in works \cite{campbell2012common, tosi2014low, liu2016fast, liu2017design, kirdoda2019geiger, liu2020ultra, fang2020ingaas}.  But in our work we fixed the total counts $R$, that can be considered as fixed $PDE$ only for low $p_{ap}$ (the SPD3 has low $p_{ap}$ and data had been approximated with exponential law quite well).  We didn't fix the $PDE$ value instead because there is not consensus for it's definition equation. On countrary, experimentally obtained counting rates can be defined only one way.

In this figure, we can see that afterpulse probability estimation $p_{ap}$ at the high values and low $\tau$ values sufficiently differs for simple and second-order models -- solid and dashed curves. However, for low $p_{ap}$ and high $\tau$ they almost coincide. We can see that using active reset has significant effects on the pap, which is especially noticeable for $PDE = 20 \ \%$. We can also see that with higher bias voltage on SPAD, which is directly related to $R$, the afterpulse probability is high.

Figure \ref{fig:apdt_spd1} b) presents the active reset pulses. The leading edge is sharp in order to remove possible triggers that may occur quickly. In this case, high-amplitude transient processes occur, but they do not influence triggers. The trailing edge is smoother to reduce the internal transients' time interval and make its amplitude lower.

\begin{figure}[h]\centering
	\includegraphics[width=0.45\linewidth]{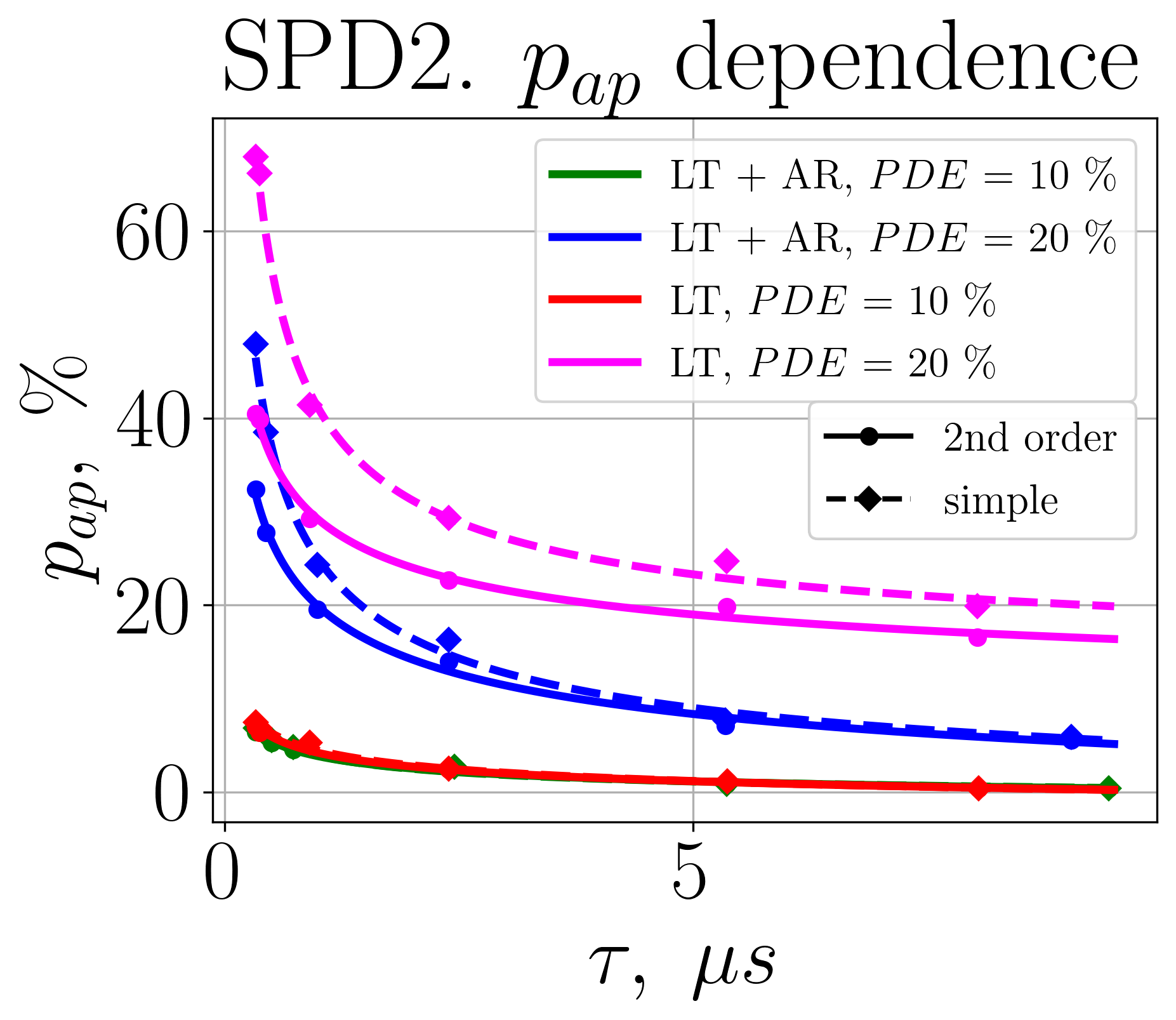} \hspace{3mm}
	\includegraphics[width=0.45\linewidth]{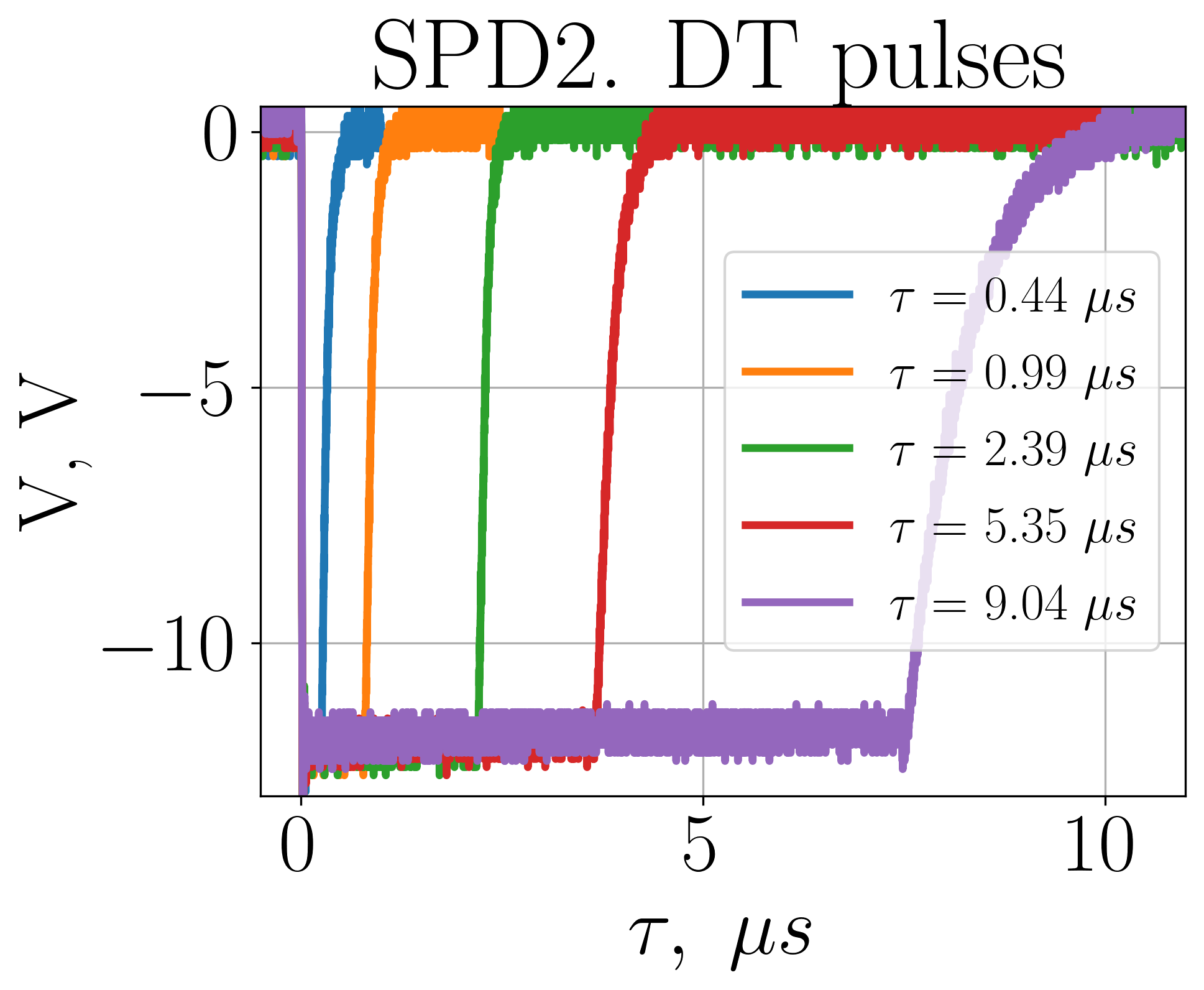}
   \caption{SPD2 data: a) Comparing $p_{ap}$ for LT and LT + AR schemes, obtained in according to simple and second-order models (diamonds and points correspondingly). Dashed lines represent the fitted curves for simple model, solid lines -- for the second-order model; b) Active reset pulse, applied to SPAD after the trigger.}
   \label{fig:apdt_spd2}
\end{figure}

Figure \ref{fig:apdt_spd2} is similar to figure \ref{fig:apdt_spd1} but performed for SPD2. We can see that for $PDE = 20 \ \%$, $p_{ap}$ is sufficiently higher than $p_{ap}$ for SPD1, and vice versa for $PDE = 10 \ \%$. This feature is not due to the control circuit but due to differences in the SPAD's characteristics. In figure b), we can see that the trailing edge is sufficiently smoother. For SPD1, that can cause intense manifestation of lowering the detector click probability due to DC bias recovering and therefore reduce SPAD efficiency for high-frequency laser pulse repetition rates.

\begin{figure}[h]\centering
	\includegraphics[width=0.45\linewidth]{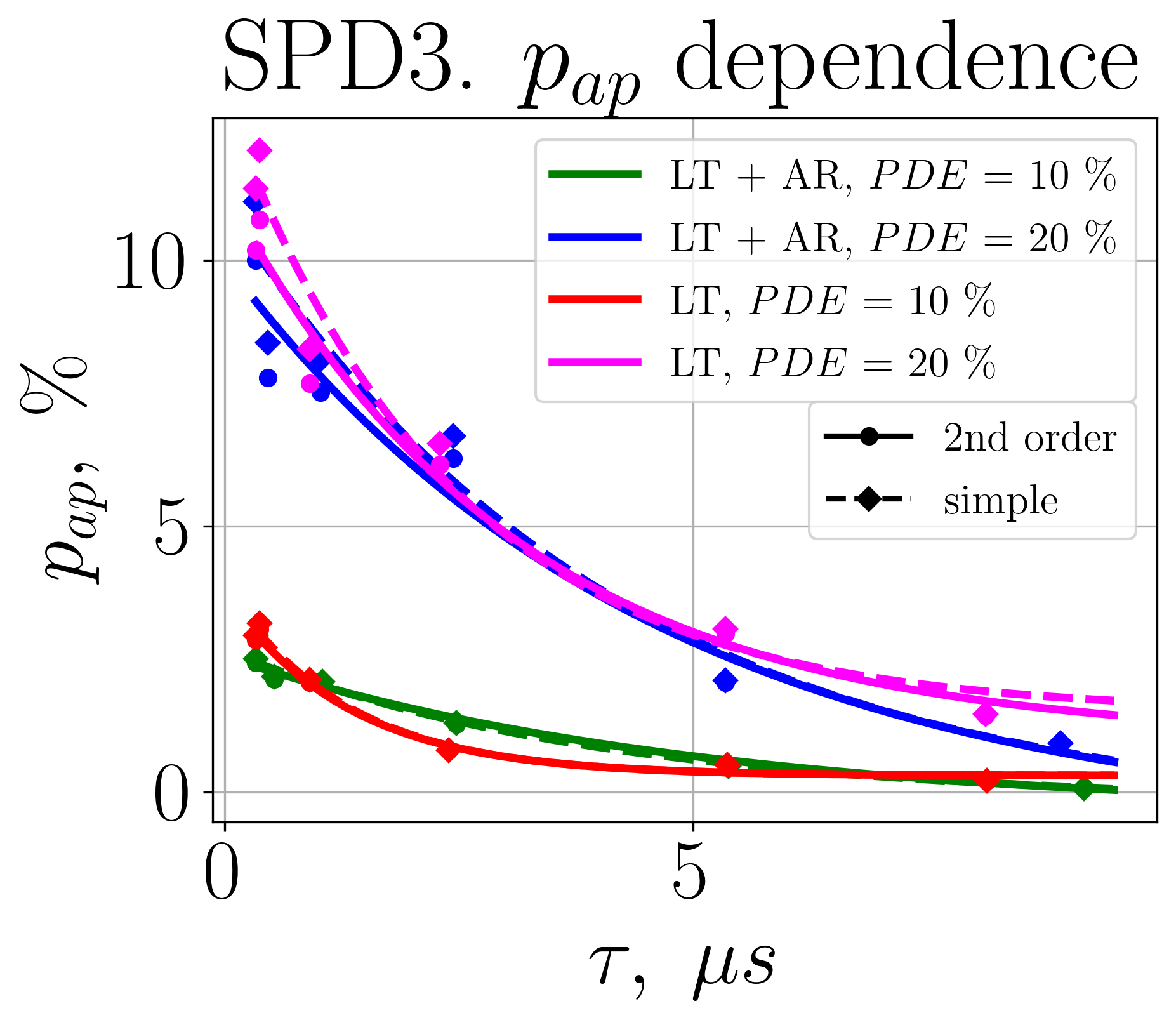} \hspace{3mm}
	\includegraphics[width=0.45\linewidth]{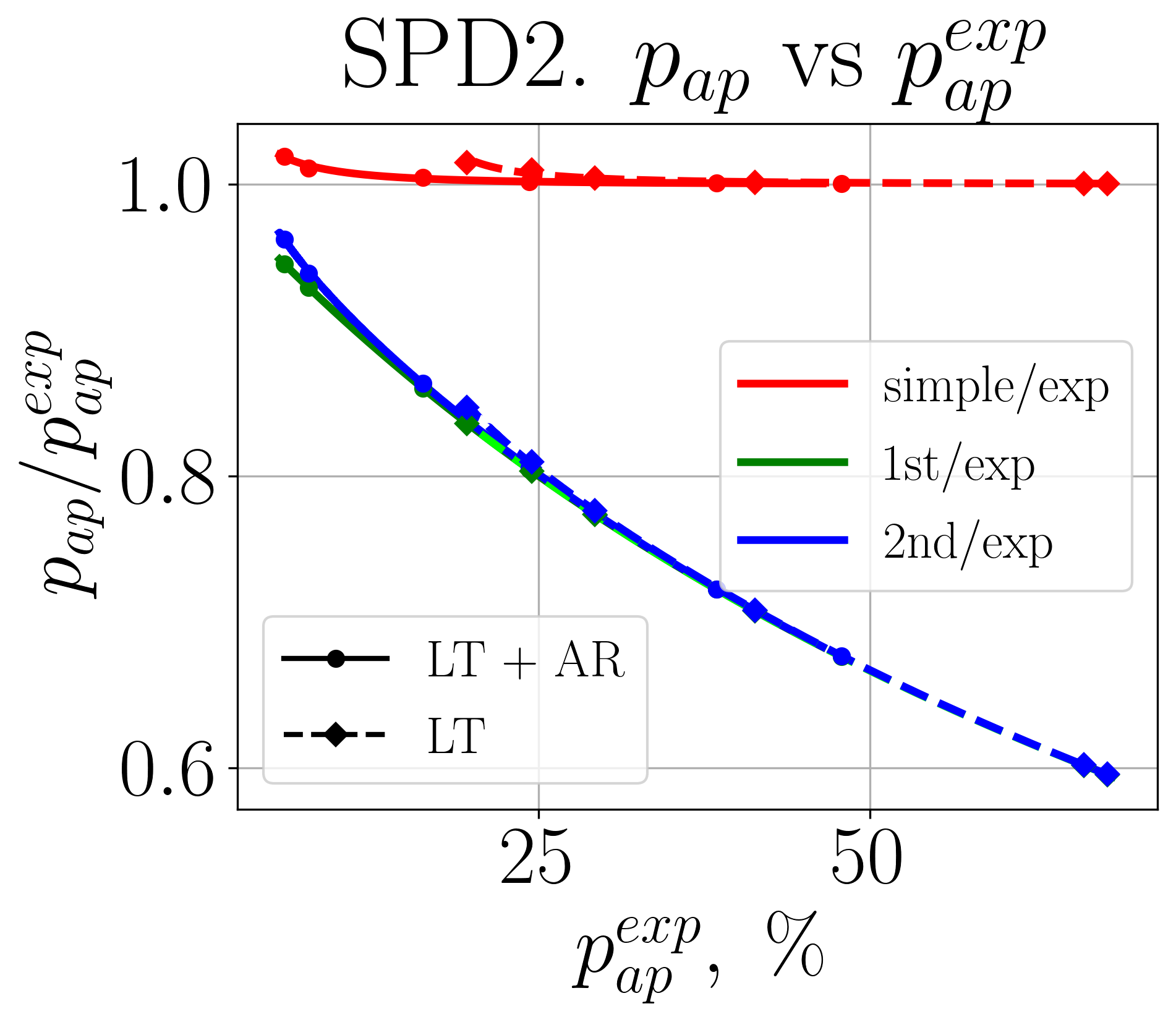}
   \caption{a) SPD3 data: comparing of $p_{ap}$ for LT and LT + AR schemes, obtained in according to simple and second-order models (diamonds and points correspondingly). Dashed lines represent the fitted curves for simple model, solid lines -- for the second-order model; b) SPD2 data: comparing of $p^{s}_{ap}$, $p^{(1)}_{ap}$ and $p^{(2)}_{ap}$ with $p^{exp}_{ap}$ for the SPD2 data and $PDE = 20 \ \%$.}
   \label{fig:apdt_spd3}
\end{figure}

Figure \ref{fig:apdt_spd3} a) presents the same, as in figures \ref{fig:apdt_spd1} and \ref{fig:apdt_spd2}, but for SPD3 data. The afterpulse probability for this detector is small enough even for low dead time and disabled active reset. For the $PDE = 10 \ \%$ data, the LT red curve lies lower than LT + AR green curve. Here it is already a matter of measurement errors and the features of the approximation of experimental data.

Figure \ref{fig:apdt_spd3} b) compares the $p^{s}_{ap}$, $p^{(1)}_{ap}$ and $p^{(2)}_{ap}$ with $p^{exp}_{ap}$ for the SPD2 data and $PDE = 20 \ \%$. We can see that the first and second-order models have a satisfactory agreement for high afterpulse probabilities and sufficiently differ from experimentally obtained $p^{exp}_{ap}$. These models differ for the afterpulse probability $p^{exp}_{ap} < 15 \ \%$. The simple model for a high $p^{exp}_{ap} > 15 \ \%$ tends to this $p^{exp}_{ap}$ value. However, for a lower $p^{exp}_{ap}$, it has even more value. There is no sufficient difference between the LT or LT + AR schemes -- they converge quite well on these graphs.

We can make the main conclusions of this schedule:
\begin{itemize}
	\item There is no difference in $p_{ap}$ for the LT and LT + AR schemes for high $\tau$ values and low $R$ (related with PDE) rates. If SPD is not designed with strict requirements for  limiting count rates and quantum efficiency, then it is permissible not to use a circuit with active reset pulses.
	\item  Choosing a model for calculating the $p_{ap}$ is essential for high afterpulse probabilities ($p^{exp}_{ap} > 5 \ \%$) -- first, or second-order models are preferable. For low values $p^{exp}_{ap}$, we can use a simple model too. According to the simple model, we get a strongly overestimated value of the afterpulse, which coincides with the experimentally found one.
\end{itemize}

\section{Conclusion}\label{sec:conclusion}

\noindent The main result of the work is that we compare schemes with latching time (LT) and with latching time and active reset (LT + AR) and its influence on the afterpulse probability. As a result of the experiments, we have shown that an active reset module could significantly reduce afterpulse probability. However, with extensive dead time ($\tau > 10 \ \mu s$) and generally low afterpulse probabilities ($p^{exp}_{ap}< 5 \ \%$), the differences between the two schemes are relatively insignificant. With low requirements for the detector ($\tau > 5 \ \mu s$ and $PDE = 10 \ \%$), the possibility of abandoning the active reset module will significantly simplify the detector's circuit design. At the same time, it slightly increases the afterpulse probability. We must mention that we performed our measurements on the $10$ kHz laser pulses repetition rate. In the LT scheme, afterpulse probability will sufficiently increase for high laser pulse repetition rates ($>1$ MHz) due to generating additional avalanche processes during the latching time. Also, we give recommendations for setting the values of the circuit dead time and latching time from the conditions of the internal transient time.

The second result is the developed approach to determining the probability of afterpulses of detectors, following three models. We have described the procedure for processing statistics from the oscilloscope histograms to obtain the experimental afterpulse probability. We introduce three models: simple, 1-st, and 2-nd order. A simple model is not appropriate for describing the afterpulsing counting statistics due to rough underlying physical processes. As a result of the experiments, we have shown that using the simple model gives rough results, which at high afterpulse probabilities simplify experimentally obtained values. We have shown that it is best to use the second-order model, which should give correct results for both small and large afterpulse probabilities. If the use of the analytical applications model is required, then the first-order model should be used because of its simple algebraic form. We have obtained that results have minor deviations from the second-order model with the afterpulse probability $p^{exp}_{ap} < 15 \ \%$, but for large values, they coincide well. We have compared our afterpulsing measurement approach with other commonly used models like Bethune method \cite{bethune2004high}, Yuan method \cite{yuan2007high}, and coincidence method \cite{zhang2009practical}. We have shown that the afterpulse probability obtained with our method is less sensitive to the laser pulses power changes. 

\bibliographystyle{IEEEtran}
\bibliography{listlit}

\begin{thebibliography}{10}
\providecommand{\url}[1]{#1}
\csname url@samestyle\endcsname
\providecommand{\newblock}{\relax}
\providecommand{\bibinfo}[2]{#2}
\providecommand{\BIBentrySTDinterwordspacing}{\spaceskip=0pt\relax}
\providecommand{\BIBentryALTinterwordstretchfactor}{4}
\providecommand{\BIBentryALTinterwordspacing}{\spaceskip=\fontdimen2\font plus
\BIBentryALTinterwordstretchfactor\fontdimen3\font minus
  \fontdimen4\font\relax}
\providecommand{\BIBforeignlanguage}[2]{{%
\expandafter\ifx\csname l@#1\endcsname\relax
\typeout{** WARNING: IEEEtran.bst: No hyphenation pattern has been}%
\typeout{** loaded for the language `#1'. Using the pattern for}%
\typeout{** the default language instead.}%
\else
\language=\csname l@#1\endcsname
\fi
#2}}
\providecommand{\BIBdecl}{\relax}
\BIBdecl

\bibitem{you2020superconducting}
L.~You, ``Superconducting nanowire single-photon detectors for quantum
  information,'' \emph{Nanophotonics}, vol.~9, no.~9, pp. 2673--2692, 2020.

\bibitem{bruschini2017ten}
C.~Bruschini, H.~Homulle, and E.~Charbon, ``Ten years of biophotonics
  single-photon {SPAD} imager applications: retrospective and outlook,'' in
  \emph{Multiphoton Microscopy in the Biomedical Sciences XVII}, vol.
  10069.\hskip 1em plus 0.5em minus 0.4em\relax International Society for
  Optics and Photonics, 2017, p. 100691S.

\bibitem{chang2021detecting}
J.~Chang, J.~Los, J.~Tenorio-Pearl, N.~Noordzij, R.~Gourgues, A.~Guardiani,
  J.~Zichi, S.~Pereira, H.~Urbach, V.~Zwiller, S.~Dorenbos, and E.~Zadeh,
  ``Detecting telecom single photons with 99.5- 2.07+ 0.5\% system detection
  efficiency and high time resolution,'' \emph{APL Photonics}, vol.~6, no.~3,
  p. 036114, 2021.

\bibitem{agnesi2020simple}
C.~Agnesi, M.~Avesani, L.~Calderaro, A.~Stanco, G.~Foletto, M.~Zahidy,
  A.~Scriminich, F.~Vedovato, G.~Vallone, and P.~Villoresi, ``Simple quantum
  key distribution with qubit-based synchronization and a self-compensating
  polarization encoder,'' \emph{Optica}, vol.~7, no.~4, pp. 284--290, 2020.

\bibitem{zhang2018experimental}
Z.~Zhang, C.~Chen, Q.~Zhuang, F.~N. Wong, and J.~H. Shapiro, ``Experimental
  quantum key distribution at 1.3 gigabit-per-second secret-key rate over a 10
  {dB} loss channel,'' \emph{Quantum Science and Technology}, vol.~3, no.~2, p.
  025007, 2018.

\bibitem{chen2020sending}
J.-P. Chen, C.~Zhang, Y.~Liu, C.~Jiang, W.~Zhang, X.-L. Hu, J.-Y. Guan, Z.-W.
  Yu, H.~Xu, J.~Lin, M.-J. Li, H.~Chen, H.~Li, L.~You, Z.~Wang, X.-B. Wang,
  Q.~Zhang, and J.-W. Pan, ``Sending-or-not-sending with independent lasers:
  Secure twin-field quantum key distribution over 509 km,'' \emph{Physical
  review letters}, vol. 124, no.~7, p. 070501, 2020.

\bibitem{kiktenko2017demonstration}
E.~O. Kiktenko, N.~O. Pozhar, A.~V. Duplinskiy, A.~A. Kanapin, A.~S. Sokolov,
  S.~S. Vorobey, A.~V. Miller, V.~E. Ustimchik, M.~N. Anufriev, A.~Trushechkin,
  R.~Yunusov, V.~Kurochkin, Y.~Kurochkin, and A.~Fedorov, ``Demonstration of a
  quantum key distribution network in urban fibre-optic communication lines,''
  \emph{Quantum Electronics}, vol.~47, no.~9, p. 798, 2017.

\bibitem{zhao2021practical}
L.-Y. Zhao, Q.-J. Wu, H.-K. Qiu, J.-L. Qian, and Z.-F. Han, ``Practical
  security of wavelength-multiplexed decoy-state quantum key distribution,''
  \emph{Physical Review A}, vol. 103, no.~2, p. 022429, 2021.

\bibitem{wang2019afterpulsing}
C.~Wang, J.~Wang, Z.~Xu, J.~Li, R.~Wang, J.~Zhao, and Y.~Wei, ``Afterpulsing
  effects in {SPAD}-based photon-counting communication system,'' \emph{Optics
  Communications}, vol. 443, pp. 202--210, 2019.

\bibitem{wang2019fast}
H.~Wang, H.~Li, L.~You, Y.~Wang, L.~Zhang, X.~Yang, W.~Zhang, Z.~Wang, and
  X.~Xie, ``Fast and high efficiency superconducting nanowire single-photon
  detector at 630 nm wavelength,'' \emph{Applied optics}, vol.~58, no.~8, pp.
  1868--1872, 2019.

\bibitem{wein2020analyzing}
S.~C. Wein, J.-W. Ji, Y.-F. Wu, F.~K. Asadi, R.~Ghobadi, and C.~Simon,
  ``Analyzing photon-count heralded entanglement generation between solid-state
  spin qubits by decomposing the master-equation dynamics,'' \emph{Physical
  Review A}, vol. 102, no.~3, p. 033701, 2020.

\bibitem{sarbazi2018impact}
E.~Sarbazi, M.~Safari, and H.~Haas, ``The impact of long dead time on the
  photocount distribution of {SPAD} receivers,'' in \emph{2018 IEEE Global
  Communications Conference (GLOBECOM)}.\hskip 1em plus 0.5em minus 0.4em\relax
  IEEE, 2018, pp. 1--6.

\bibitem{smirnov2018sequences}
M.~A. Smirnov, N.~S. Perminov, R.~R. Nigmatullin, A.~A. Talipov, and S.~A.
  Moiseev, ``Sequences of the ranged amplitudes as a universal method for fast
  noninvasive characterization of {SPAD} dark counts,'' \emph{Applied optics},
  vol.~57, no.~1, pp. 57--61, 2018.

\bibitem{kramnik2020efficient}
D.~Kramnik and R.~J. Ram, ``Efficient statistical separation of primary dark
  counts and afterpulses in free-running {SPAD}s,'' in \emph{CLEO: Applications
  and Technology}.\hskip 1em plus 0.5em minus 0.4em\relax Optical Society of
  America, 2020, pp. JTh2A--29.

\bibitem{fan2020optimizing}
G.-J. Fan-Yuan, J.~Teng, S.~Wang, Z.-Q. Yin, W.~Chen, D.-Y. He, G.-C. Guo, and
  Z.-F. Han, ``Optimizing single-photon avalanche photodiodes for dynamic
  quantum key distribution networks,'' \emph{Physical Review Applied}, vol.~13,
  no.~5, p. 054027, 2020.

\bibitem{owens1994photon}
P.~Owens, J.~Rarity, P.~Tapster, D.~Knight, and P.~Townsend, ``Photon counting
  with passively quenched germanium avalanche,'' \emph{Applied Optics},
  vol.~33, no.~30, pp. 6895--6901, 1994.

\bibitem{wang2016non}
F.-X. Wang, W.~Chen, Y.-P. Li, D.-Y. He, C.~Wang, Y.-G. Han, S.~Wang, Z.-Q.
  Yin, and Z.-F. Han, ``Non-markovian property of afterpulsing effect in
  single-photon avalanche detector,'' \emph{Journal of Lightwave Technology},
  vol.~34, no.~15, pp. 3610--3615, 2016.

\bibitem{kang2003afterpulsing}
Y.~Kang, D.~Bethune, W.~Risk, and Y.-H. Lo, ``Afterpulsing of single-photon
  avalanche photodetectors,'' in \emph{The 16th Annual Meeting of the IEEE
  Lasers and Electro-Optics Society, 2003. LEOS 2003.}, vol.~2.\hskip 1em plus
  0.5em minus 0.4em\relax IEEE, 2003, pp. 775--776.

\bibitem{bethune2004high}
D.~S. Bethune, W.~P. Risk, and G.~W. Pabst, ``A high-performance integrated
  single-photon detector for telecom wavelengths,'' \emph{Journal of modern
  optics}, vol.~51, no. 9-10, pp. 1359--1368, 2004.

\bibitem{yuan2007high}
Z.~Yuan, B.~Kardynal, A.~Sharpe, and A.~Shields, ``High speed single photon
  detection in the near infrared,'' \emph{Applied Physics Letters}, vol.~91,
  no.~4, p. 041114, 2007.

\bibitem{namekata20091}
N.~Namekata, S.~Adachi, and S.~Inoue, ``1.5 {GHz} single-photon detection at
  telecommunication wavelengths using sinusoidally gated {InGaAs/InP} avalanche
  photodiode,'' \emph{Optics express}, vol.~17, no.~8, pp. 6275--6282, 2009.

\bibitem{nambu2011efficient}
Y.~Nambu, S.~Takahashi, K.~Yoshino, A.~Tanaka, M.~Fujiwara, M.~Sasaki,
  A.~Tajima, S.~Yorozu, and A.~Tomita, ``Efficient and low-noise single-photon
  avalanche photodiode for 1.244-{GH}z clocked quantum key distribution,''
  \emph{Optics express}, vol.~19, no.~21, pp. 20\,531--20\,541, 2011.

\bibitem{zhang2009practical}
J.~Zhang, R.~Thew, C.~Barreiro, and H.~Zbinden, ``Practical fast gate rate
  {InGaAs/InP} single-photon avalanche photodiodes,'' \emph{Applied Physics
  Letters}, vol.~95, no.~9, p. 091103, 2009.

\bibitem{zhang20102}
J.~Zhang, P.~Eraerds, N.~Walenta, C.~Barreiro, R.~Thew, and H.~Zbinden, ``2.23
  {GHz} gating {InGaAs/InP} single-photon avalanche diode for quantum key
  distribution,'' in \emph{Advanced Photon Counting Techniques IV}, vol.
  7681.\hskip 1em plus 0.5em minus 0.4em\relax International Society for Optics
  and Photonics, 2010, p. 76810Z.

\bibitem{zhang2014electro}
Y.~Zhang, X.~Zhang, Y.~Shi, Z.~Ying, and S.~Wang, ``Electro-optic modulator
  based gate transient suppression for sine-wave gated {InGaAs/InP} single
  photon avalanche photodiode,'' \emph{Optical Engineering}, vol.~53, no.~6, p.
  067102, 2014.

\bibitem{restelli2012time}
A.~Restelli, J.~C. Bienfang, and A.~L. Migdall, ``Time-domain measurements of
  afterpulsing in {InGaAs/InP} {SPAD} gated with sub-nanosecond pulses,''
  \emph{Journal of Modern Optics}, vol.~59, no.~17, pp. 1465--1471, 2012.

\bibitem{itzler2011advances}
M.~A. Itzler, X.~Jiang, M.~Entwistle, K.~Slomkowski, A.~Tosi, F.~Acerbi,
  F.~Zappa, and S.~Cova, ``Advances in {InGaAsP}-based avalanche diode single
  photon detectors,'' \emph{Journal of Modern Optics}, vol.~58, no. 3-4, pp.
  174--200, 2011.

\bibitem{zhang2009comprehensive}
J.~Zhang, R.~Thew, J.-D. Gautier, N.~Gisin, and H.~Zbinden, ``Comprehensive
  characterization of {InGaAs--InP} avalanche photodiodes at 1550 nm with an
  active quenching asic,'' \emph{IEEE Journal of Quantum Electronics}, vol.~45,
  no.~7, pp. 792--799, 2009.

\bibitem{arahira2016effects}
S.~Arahira and H.~Murai, ``Effects of afterpulse events on performance of
  entanglement-based quantum key distribution system,'' \emph{Japanese Journal
  of Applied Physics}, vol.~55, no.~3, p. 032801, 2016.

\bibitem{Klyshko_1980}
D.~N. Klyshko, ``Use of two-photon light for absolute calibration of
  photoelectric detectors,'' \emph{Soviet Journal of Quantum Electronics},
  vol.~10, no.~9, pp. 1112--1117, Sep 1980.

\bibitem{kwiat1994absolute}
P.~G. Kwiat, A.~M. Steinberg, R.~Y. Chiao, P.~H. Eberhard, and M.~D. Petroff,
  ``Absolute efficiency and time-response measurement of single-photon
  detectors,'' \emph{Applied optics}, vol.~33, no.~10, pp. 1844--1853, 1994.

\bibitem{brida2000quantum}
G.~Brida, S.~Castelletto, I.~P. Degiovanni, C.~Novero, and M.~L. Rastello,
  ``Quantum efficiency and dead time of single-photon counting photodiodes: a
  comparison between two measurement techniques,'' \emph{Metrologia}, vol.~37,
  no.~5, p. 625, 2000.

\bibitem{polyakov2007high}
S.~V. Polyakov and A.~L. Migdall, ``High accuracy verification of a
  correlated-photon-based method for determining photon-counting detection
  efficiency,'' \emph{Optics Express}, vol.~15, no.~4, pp. 1390--1407, 2007.

\bibitem{tosi2014low}
A.~Tosi, N.~Calandri, M.~Sanzaro, and F.~Acerbi, ``Low-noise, low-jitter, high
  detection efficiency {InGaAs/InP} single-photon avalanche diode,'' \emph{IEEE
  Journal of selected topics in quantum electronics}, vol.~20, no.~6, pp.
  192--197, 2014.

\bibitem{liu2020ultra}
J.~Liu, Y.~Xu, Y.~Li, Y.~Gu, Z.~Liu, and X.~Zhao, ``Ultra-low dead time
  free-running {InGaAsP} single-photon detector with active quenching,''
  \emph{Journal of Modern Optics}, vol.~67, no.~13, pp. 1184--1189, 2020.

\bibitem{liu2021exploiting}
J.~Liu, Y.~Xu, Y.~Li, Z.~Liu, and X.~Zhao, ``Exploiting the single-photon
  detection performance of {InGaAs} negative-feedback avalanche diode with fast
  active quenching,'' \emph{Optics Express}, vol.~29, no.~7, pp.
  10\,150--10\,161, 2021.

\bibitem{korzh2015afterpulsing}
B.~Korzh, T.~Lunghi, K.~Kuzmenko, G.~Boso, and H.~Zbinden, ``Afterpulsing
  studies of low-noise {InGaAs/InP} single-photon negative-feedback avalanche
  diodes,'' \emph{Journal of Modern Optics}, vol.~62, no.~14, pp. 1151--1157,
  2015.

\bibitem{humer2015simple}
G.~Humer, M.~Peev, C.~Schaeff, S.~Ramelow, M.~Stip{\v{c}}evi{\'c}, and
  R.~Ursin, ``A simple and robust method for estimating afterpulsing in single
  photon detectors,'' \emph{Journal of Lightwave Technology}, vol.~33, no.~14,
  pp. 3098--3107, 2015.

\bibitem{itzler2012power}
M.~A. Itzler, X.~Jiang, and M.~Entwistle, ``Power law temporal dependence of
  {InGaAs/InP SPAD} afterpulsing,'' \emph{Journal of Modern Optics}, vol.~59,
  no.~17, pp. 1472--1480, 2012.

\bibitem{horoshko2017afterpulsing}
D.~Horoshko, V.~Chizhevsky, and S.~Y. Kilin, ``Afterpulsing model based on the
  quasi-continuous distribution of deep levels in single-photon avalanche
  diodes,'' \emph{Journal of Modern Optics}, vol.~64, no.~2, pp. 191--195,
  2017.

\bibitem{ziarkash2018comparative}
A.~W. Ziarkash, S.~K. Joshi, M.~Stip{\v{c}}evi{\'c}, and R.~Ursin,
  ``Comparative study of afterpulsing behavior and models in single photon
  counting avalanche photo diode detectors,'' \emph{Scientific reports},
  vol.~8, no.~1, pp. 1--8, 2018.

\bibitem{koziy2021investigating}
A.~Koziy, A.~Tayduganov, A.~Losev, V.~Zavodilenko, A.~Gorbatsevich, and
  Y.~Kurochkin, ``Investigating the coherent state detection probability of
  {InGaAs/InP} {SPAD}-based single-photon detectors,'' 2021.

\bibitem{campbell2012common}
J.~C. Campbell, W.~Sun, Z.~Lu, M.~A. Itzler, and X.~Jiang, ``Common-mode
  cancellation in sinusoidal gating with balanced {InGaAs/InP} single photon
  avalanche diodes,'' \emph{IEEE Journal of Quantum Electronics}, vol.~48,
  no.~12, pp. 1505--1511, 2012.

\bibitem{liu2016fast}
J.~Liu, Y.~Li, L.~Ding, Y.~Wang, T.~Zhang, Q.~Wang, and J.~Fang, ``Fast
  active-quenching circuit for free-running {InGaAs (P)/InP} single-photon
  avalanche diodes,'' \emph{IEEE Journal of Quantum Electronics}, vol.~52,
  no.~10, pp. 1--6, 2016.

\bibitem{liu2017design}
J.~Liu, T.~Zhang, Y.~Li, L.~Ding, J.~Tao, Y.~Wang, Q.~Wang, and J.~Fang,
  ``Design and characterization of free-running {InGaAsP} single-photon
  detector with active-quenching technique,'' \emph{Journal of Applied
  Physics}, vol. 122, no.~1, p. 013104, 2017.

\bibitem{kirdoda2019geiger}
J.~Kirdoda, D.~C. Dumas, R.~W. Millar, M.~M. Mirza, D.~J. Paul, K.~Kuzmenko,
  P.~Vines, Z.~Greener, and G.~S. Buller, ``Geiger mode {Ge-on-Si}
  single-photon avalanche diode detectors,'' in \emph{2019 IEEE 2nd British and
  Irish Conference on Optics and Photonics (BICOP)}.\hskip 1em plus 0.5em minus
  0.4em\relax IEEE, 2019, pp. 1--4.

\bibitem{fang2020ingaas}
Y.-Q. Fang, W.~Chen, T.-H. Ao, C.~Liu, L.~Wang, X.-J. Gao, J.~Zhang, and J.-W.
  Pan, ``{InGaAs/InP} single-photon detectors with 60\% detection efficiency at
  1550 nm,'' \emph{Review of Scientific Instruments}, vol.~91, no.~8, p.
  083102, 2020.

\end{thebibliography}
%\bibliography{iter8.bbl}

\end{document}